\documentclass[structabstract]{aa}  
\pdfoutput=1 
\usepackage{natbib, graphicx}
\usepackage{amsmath}
\usepackage[flushleft]{threeparttable}
\usepackage{subfigure}
\usepackage{color}
\usepackage{txfonts}
\usepackage{multirow}
\usepackage{hyperref}
\usepackage{booktabs}
\usepackage{pdflscape}
\usepackage{adjustbox}
\usepackage{caption}

\DeclareCaptionLabelFormat{continued}{#1~#2~Continued}
\begin{document}

   \title{Search for lithium-rich giants in 32 open clusters with high-resolution spectroscopy\thanks{Based on observations collected at the La Silla Observatory,
ESO (Chile), with HARPS 3.6\,m (runs 075.C-0140, 076.C-0429, 078.C-0133, 079.C-0329, 080.C-0071, 081.C-0119, 082.C-0333, 083.C-0413, 091.C-0438, 092.C-0282, 094.C-0297, 099.C-0304, 0100.C-0888, 0101.C-0274, and 0102.C-0812) and with UVES/VLT at the Cerro Paranal Observatory (run 079.C-0131).}}
   \author{M. Tsantaki\inst{1,2} 
          \and    
	  E. Delgado-Mena\inst{1}
          \and    
	  D. Bossini\inst{1}
         \and S. G. Sousa\inst{1} 
         \and E. Pancino\inst{2} 
         \and J. H. C. Martins\inst{1}}
\institute{Instituto de Astrof\'{i}sica e Ci\^encias do Espaço, Universidade do Porto, CAUP, Rua das Estrelas, 4150-762 Porto, Portugal \\
    \email{maria.tsantaki@inaf.it} 
    \and {INAF -- Osservatorio Astrofisico di Arcetri, Largo E. Fermi 5, 50125 Firenze, Italy}}
        
   \date{Received XXXX; accepted XXXX}
   \authorrunning{M. Tsantaki}
   \titlerunning{Li-rich giant stars in open clusters}
   \abstract
{Lithium-rich giant stars are rare and their existence challenges our understanding of stellar structure and evolution. In particular, open clusters constrain well the mass and age of their members, and therefore, offer a unique opportunity to understand the evolutionary stage where Li enrichment occurs.}
{We profit from the high-quality sample gathered to search for planets in open clusters with HARPS and UVES, in order to search for Li-rich giants and to identify the Li enrichment mechanisms responsible.}
{We derive stellar parameters for 247 stars belonging to 32 open clusters, with 0.07\,Ga\,$<$\,ages\,$<$\,3.6\,Ga. We employed the spectral synthesis technique code FASMA for the abundance analysis of 228 stars from our sample. We also determined ages, distances, and extinction using astrometry and photometry from {\em Gaia} and PARSEC isochrones to constrain their evolutionary stage. Our sample covers a wide range of stellar masses from 1 to more than 6\,M$_{\odot}$ where the majority of the masses are above 2\,M$_{\odot}$.}
{We have found 14 canonical Li-rich giant stars which have experienced the first dredge-up. This corresponds to 6\% of our total sample, which is higher than what is typically found for field stars. The majority of the stars (11/14) are located at the red clump, two lie on the red giant branch, and for one we could not conclude on its evolutionary stage. Apart from the canonical limit, we use the maximum Li abundance of the progenitor stars as a criterion for Li enrichment and find 12 Li enriched stars (5/12 appear in the red clump, 5/12 at the upper red giant branch and two we could not conclude on its evolutionary stage). We find Li enhancement also among eight stars which have passed the first dredge up and show strong Li lines based on the fact that stars at the same evolutionary stage in the same cluster have significantly different Li abundances. We confirm that giants with higher Li abundance correspond to a higher fraction of fast-rotating giants, suggesting a connection between Li enhancement and stellar rotation as predicted by stellar models.} 
{Our Li-rich giants are found in various evolutionary stages implying that no unique Li production mechanism is responsible for Li enrichment but rather different intrinsic or external mechanisms can be simultaneously at play.}
\keywords{stars: abundances – planetary systems – stars: rotation – stars: evolution – techniques: spectroscopic}

\maketitle

\section{Introduction}\label{intro}

The abundance of lithium (Li) in stellar atmospheres suffers dramatic changes throughout the lifetime of a star in contrast to heavier elements which remain approximately constant. Lithium is therefore, a sensitive indicator of stellar evolution.

The vast majority of stars with mass similar to the Sun are expected to only destroy lithium over the course of their lives via proton capture reactions. The interior structure of solar main-sequence stars is comprised of a radiative core and a quite shallow convective envelope which protects the surface Li from exposure to temperatures high enough to destroy it ($\sim$2.6\,$\times$\,10$^{6}$\,K). According to the stellar evolution theory, when a star leaves the main sequence, the depth of the convective envelope increases as it ascends towards the red giant branch (RGB). As the convective envelope expands significantly deeper, the surviving lithium gets substantially diluted. At this stage, the star experiences the first dredge-up (FDU) where hot material from the deep internal layers is mixed towards the surface \citep{Iben1967a, Iben1967b} and it is hot enough that Li is destroyed. The FDU, therefore, causes dilution of the surface Li to abundances typically below 1.5\,dex (see Sect.~\ref{section_results} for a discussion on the canonical limit of Li-rich giants) which is the post dredge-up abundance expectation for Population I stars with $\sim$\,1\,M$_{\odot}$ \citep{Lambert1980, Brown1989, Mallik1999, Charbonnel2000}. 

The unusual discovery of a Li-rich giant with a post-FDU abundance A(Li)\,=\,3.2\,dex\footnote{Lithium abundance is expressed as: A(Li) = $\log$[N(Li)/N(H)] + 12, where N(Li) and N(H) are the number densities of Li and H atoms.} was the first case to be studied  \citep{Wallerstein1982}. Since then, the discoveries of Li-rich giants have increased remarkably mainly due to the large spectroscopic surveys, such as the {\em Gaia}-ESO, GALAH, and LAMOST \citep[e.g.][]{Casey2016, Deepak2019, Gao2019, Singh2019}. Li-rich giants are still rare in the field, representing 1-2\% of giants \citep[e.g.][]{Deepak2019} and even rarer in globular clusters \citep[e.g.][]{Sanna2020} challenging the standard stellar evolution models.

Several scenarios have been proposed to explain the unexpectedly high Li abundances of giant stars. On one hand, an external contamination might increase the Li content in the photosphere caused by \textit{i)} accretion of planetary material \citep{Alexander1967, Siess1999, Adamow2012, Aguilera2016, Delgado2016}, \textit{ii)} contamination by a companion located at the asymptotic giant branch (AGB) that has already been through a Li-production process \citep[e.g.][]{Sackmann1999}, \textit{iii)} tidal spin-up from a binary companion \citep[e.g.][]{Casey2019}, \textit{iv)} or by accretion of interstellar material enriched by supernova \citep[e.g.][]{Woosley1995}. On the other hand, internal Li production is possible through the Cameron-Fowler mechanism \citep[CF,][]{Cameron1971}. The stellar interior reaches high enough temperatures to produce the Li predecessor, beryllium (Be), and then, Li is created by the Be decay. After the  $^{3}$He($\alpha$,$\gamma$)$^{7}$Be reaction occurs, Be has to be rapidly transported to the outer cooler regions of the stellar surface in order to decay to Li where it can survive. For the AGB stars, the CF mechanism can be responsible for the Li enrichment because the temperature at the bottom of the convective envelope is high enough to produce Be, referred as the hot bottom burning (HBB) process for stars with masses higher than 4\,M$_{\odot}$. However, in less evolved RGB stars, the lower envelope is too cool to produce Be. In this case, an extra mixing is required for the low-mass stars to convey material through the outer and inner layers where in the latter the environment is hot enough to produce Be. The mechanisms for the extra mixing which can produce Li proposed in the literature are mainly the thermohaline mixing \citep[e.g.][]{Eggleton2008}, rotation-induced mixing \citep[e.g.][]{Denissenkov2004}, and magnetic buoyancy \citep[e.g.][]{Nordhaus2008}. 

The evolutionary phases which have caught attention to trigger internal Li enrichment are: giants at the luminosity bump (LB) on the RGB and at the RGB tip of low-mass giants (M$\lesssim$\,2.2\,M$_{\odot}$) which experience a He-core flash before entering the red clump (RC) phase. At the LB, the outward hydrogen burning shell reaches the mean molecular weight discontinuity caused by the FDU where it is burnt at a lower rate, leading to a slight decrease in luminosity \citep[e.g.][]{Christensen2015}. This phenomenon causes an extra mixing to trigger the CF mechanism and produce extra Li \citep{Charbonnel2000, Eggleton2008}. The decrease in luminosity at the LB is represented as a loop at the Hertzsprung-Russell (HR) diagram. 

Furthermore, recent studies have observationally found that Li-rich low mass giants (M$\lesssim$\,2\,M$_{\odot}$) are almost exclusively RC stars \citep[e.g.][]{Singh2019, Deepak2021, Yan2021}. The mechanisms proposed to explain Li enrichment at this evolutionary phase are still under investigation and focus on \textit{i)} models which produce Li at the RGB tip, before the onset of He flash, via the inclusion of the neutrino magnetic moment \citep{Mori2021} or \textit{ii)} models which produce Li during the main He-core flash, by an ad-hoc mixing from internal gravity waves \citep{Schwab2020}. This ubiquitous Li production at the RC however, could be due to population effects of field stars as mass and chemical composition are important factors for the effectiveness of Li destruction \citep{Magrini2021, Chaname2021}. Unlike field populations, the giants in a stellar cluster, span a narrow range of stellar mass and composition which makes clusters ideal environments to study Li enhancement. The key to understand Li enrichment lies in finding the evolutionary phase of Li-rich giants. In this work, we study the Li content of a large sample of giant stars in open clusters (OCs) where we can well constrain their evolutionary stage. 

The paper is organised as follows: in Sects.~\ref{method}-\ref{Li_determination} we present the data and explain the spectroscopic analysis to obtain the stellar parameters, and in Sect.~\ref{section_results} we discuss the results. The conclusions are presented in Sect.~\ref{conclusions}.

\section{Observations and stellar parameters}\label{method}

\begin{figure}
  \centering
  \includegraphics[width=1.0\linewidth]{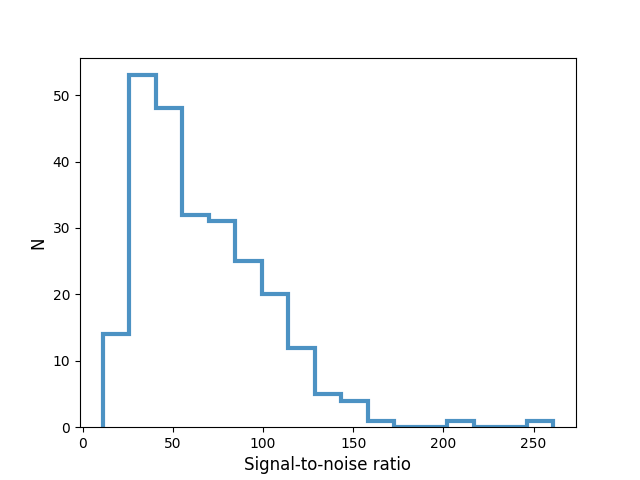}    
  \caption{The distribution of the S/N of the combined spectra in our whole sample.}
  \label{snr}
\end{figure}

\begin{figure}
 \centering
 \includegraphics[width=1.0\linewidth]{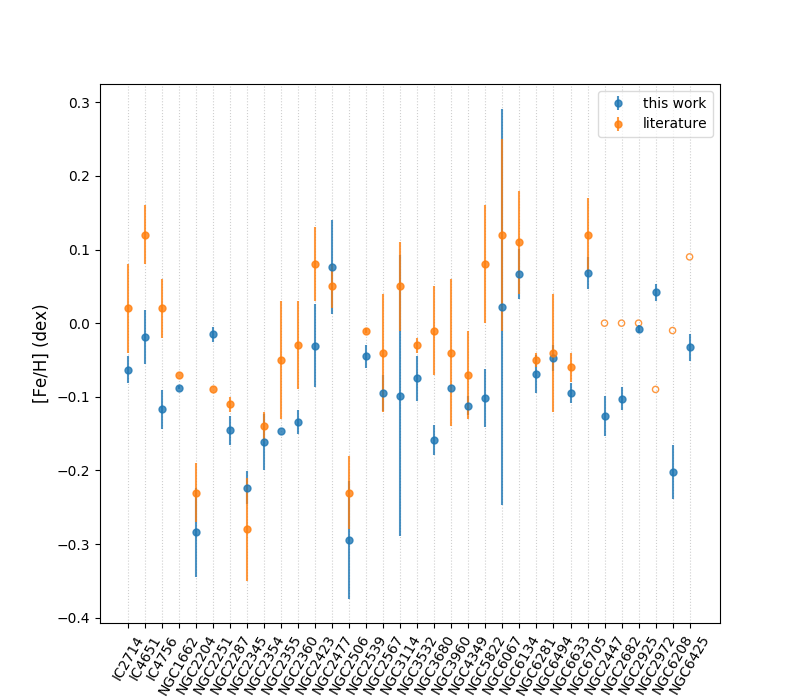}    
 \caption{The average iron metallicity derived for each cluster. Open symbols indicate non spectroscopic literature values.}
 \label{feh_cluster}
\end{figure}

\begin{figure}
\centering
\includegraphics[width=1.0\linewidth]{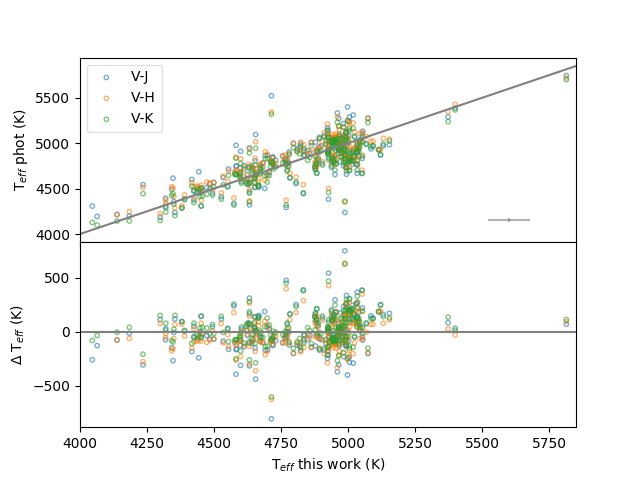}    
\caption{Spectroscopic T$_{\rm eff}$ from this work in comparison with photometric T$_{\rm eff}$ derived from different colour calibrations, (V--\textit{J}) in blue, (V--\textit{H}) in orange, (V--\textit{K$_{s}$}) in green. $\Delta$ T$_{\rm eff}$ in the bottom plot represents spectroscopic minus photometric values.}
\label{teff_cluster}
\end{figure}

\begin{figure}
\centering
\includegraphics[width=1.0\linewidth]{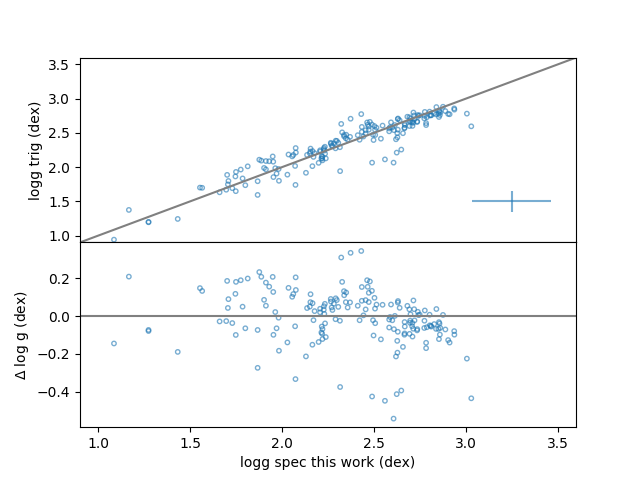}    
\caption{Spectroscopic $\log g$ from this work in comparison with trigonometric $\log g$. $\Delta$ $\log g$ in the bottom plot represents spectroscopic minus photometric values.}
\label{logg_cluster}
\end{figure}

\begin{table}
\centering
\caption{Mean iron metallicity and standard deviation ($\sigma$) for the 34 clusters.}
\label{feh_cluster_table}
\begin{tabular}{lccccc}
\hline\hline
\multirow{2}{*}{Clusters} & $[Fe/H]$ & $\sigma$ & \multirow{2}{*}{N} & $[Fe/H]_{lit}$ & \multirow{2}{*}{Reference} \\
                          & (dex)    & (dex)    &  &  (dex)   \\
  \hline
IC2714   & --0.06 & 0.02 & 8  &  0.02 $\pm$ 0.06    & (1) \\
IC4651   & --0.01 & 0.04 & 11 &  0.12 $\pm$ 0.04    & (1) \\
IC4756   & --0.12 & 0.03 & 13 &  0.02 $\pm$ 0.04    & (1) \\
NGC1662  & --0.09 & 0.00 & 2  & --0.07              & (2) \\
NGC2204  & --0.28 & 0.06 & 2  &  --0.23 $\pm$ 0.04  & (7) \\
NGC2251  & --0.01 & 0.01 & 3  & --0.09        & (1) \\
NGC2287  & --0.14 & 0.02 & 4  &  --0.11 $\pm$ 0.01  & (1) \\
NGC2345  & --0.22 & 0.02 & 4  &  --0.28 $\pm$ 0.07  & (3) \\
NGC2354  & --0.17 & 0.04 & 6  &  --0.14 $\pm$ 0.02  & (2) \\
NGC2355  & --0.15 &      & 1  &  --0.05 $\pm$ 0.08  & (1) \\
NGC2360  & --0.13 & 0.02 & 8  &  --0.03 $\pm$ 0.06  & (1) \\
NGC2423  & --0.03 & 0.06 & 4  &  0.08 $\pm$ 0.05    & (1) \\
NGC2447	 & --0.13 & 0.03 & 3  & 0.00                & (4) \\
NGC2477  & 0.08   & 0.06 & 36 & 0.07 $\pm$ 0.03     & (1) \\
NGC2506  & --0.29 & 0.08 & 3  &  --0.23 $\pm$ 0.05  & (1) \\
NGC2539  & --0.04 & 0.02 & 9  &  --0.01 $\pm$ 0.004 & (2) \\
NGC2567  & --0.09 & 0.02 & 2  &  --0.04 $\pm$ 0.08  & (1) \\
NGC2682  & --0.10 & 0.02 & 3  &  0.00               & (4) \\
NGC2925  & --0.01 & 0.01 & 2  & 0.00                & (4)\\
NGC2972  & 0.04   & 0.01 & 2  & --0.09              & (4) \\
NGC3114  & --0.10 & 0.19 & 8  &  0.05 $\pm$ 0.06    & (1) \\
NGC3532  & --0.08 & 0.03 & 7  &  --0.03 $\pm$ 0.01  & (2) \\
NGC3680  & --0.15 & 0.02 & 5  &  --0.01 $\pm$ 0.06  & (1) \\
NGC3960  & --0.09 &      & 1  &  --0.04 $\pm$ 0.10  & (1) \\
NGC4349  & --0.11 & 0.01 & 2  &  --0.07 $\pm$ 0.06  & (1) \\
NGC5822  & --0.10 & 0.04 & 12 &  0.08 $\pm$ 0.08    & (1) \\
NGC6067  & 0.02   & 0.26 & 3  &  0.12 $\pm$ 0.13    & (1) \\
NGC6134  & 0.07   & 0.03 & 3  &  0.11 $\pm$ 0.07    & (1) \\
NGC6208  & --0.20 & 0.04 & 2  & --0.01              & (4) \\
NGC6281  & --0.07 & 0.03 & 2  &  --0.05 $\pm$ 0.01  & (2)\\
NGC6425  & --0.03 & 0.02 & 2  & 0.09                & (4) \\
NGC6494  & --0.05 & 0.02 & 4  &  --0.04 $\pm$ 0.08  & (1) \\
NGC6633  & --0.09 & 0.01 & 3  &  --0.06 $\pm$ 0.02  & (5) \\
NGC6705  & 0.07   & 0.02 & 29 &  0.12 $\pm$ 0.05    & (6) \\
\hline
\end{tabular} 
\begin{tablenotes}
\item N represents the number of stars in each cluster excluding non members. The literature iron metallicity and their references are shown in the last columns. References: (1) \cite{Heiter2014}, (2) \cite{Casamiquela2021}, (3) \cite{Alonso2019}, (4) \cite{Bossini2019}, (5) \cite{Spina2017}, (6) \cite{Magrini2014}, (7) \cite{Jacobson2011}.
\end{tablenotes}
\end{table} 

\begin{table}
\centering
\caption{Statistics from the comparison between the parameters from HARPS and UVES spectra.}
\label{uves_harps}
\begin{tabular}{lccccc}
\hline\hline
$\Delta$                   & Mean   & Median & $\sigma$ & MAD & N \\
\hline
$\Delta$ T$_{\rm eff}$ (K) & --2    & --1    & 13   & 10   & 43 \\
$\Delta$ $\log g$ (dex)    & --0.13 & --0.13 & 0.08 & 0.06 & 43 \\
$\Delta$ [Fe/H] (dex)      & --0.01 & --0.01 & 0.03 & 0.02 & 43 \\
\hline
\end{tabular}
\end{table} 

\begin{table}
\centering
\caption{Statistics from the comparison between the parameters of this work and the T$_{\rm eff}$ from photometry and trigonometric $\log g$.}
\label{stellar_params_diff_table}
\scalebox{0.97}{
\begin{tabular}{lccccc}
\hline\hline
$\Delta$                     & Mean   & Median & $\sigma$ & MAD & N \\
\hline
$\Delta$ [Fe/H] (dex)                                    & 0.06   & 0.05   & 0.06 & 0.03 & 28 \\
$\Delta$ T$^{photo}_{\rm eff}$ (V--\textit{J}) (K)       & 2      & --14   & 178  & 106  & 219 \\
$\Delta$ T$^{photo}_{\rm eff}$ (V--\textit{H}) (K)       & --13   & --38   & 146  & 85   & 217 \\
$\Delta$ T$^{photo}_{\rm eff}$ (V--\textit{K$_{s}$}) (K) & 30     & 22     & 145  & 97   & 219 \\
$\Delta$ T$_{\rm eff}$ (K)                       & 7      & --11   & 155  & 94   & 219 \\
$\Delta$ $\log g$ (dex)                                  & --0.01 & --0.02 & 0.13 & 0.07 & 208 \\
\hline
\end{tabular}}
\end{table} 

The baseline sample of this work is taken from \citet[][hereafter DM16]{Delgado2016} who used data from the planet search surveys described in \cite{Lovis2007} and in the subsequent study of \cite{Delgado2018}. In addition, we collected spectra from other programs in the ESO archive, and new spectra from the above planet search surveys. Our total sample is composed of 250 evolved stars in 34 different open clusters, all of them observed with the HARPS spectrograph \citep{Mayor2003} at the ESO 3.6\,m telescope (R\,$\sim$\,115\,000) in La Silla (Chile). The individual spectra of each star were reduced using the HARPS pipeline, corrected for their radial velocity (RV) shift, and finally, combined with IRAF tools \citep{iraf1993} to increase the signal-to-noise ratio (S/N). We also include 53 stars observed with UVES spectrograph (R\,$\sim$\,50\,000) at the VLT in Paranal (Chile) from DM16 where 43 of them have also HARPS spectra in our sample and ten have only UVES spectra. To maximize homogeneity, we used only the HARPS spectra except for the ten targets which do not have HARPS observations available and for which we used UVES observations. The rest of the UVES spectra are used for comparison tests to validate homogeneity in our analysis method (see below). The coordinates and magnitudes of the sample are shown in Table~\ref{sample_params} and the distribution of the S/N\footnote{The S/N is calculated by the PyAstronomy function, \texttt{estimateSNR}, using the median value from small intervals.} is plotted in Fig.~\ref{snr}.

The stellar parameters (effective temperature, T$_{\rm eff}$, surface gravity, $\log g$, iron metallicity, $[Fe/H]$, and projected rotational velocity, $\upsilon \sin i$), are determined in a homogeneous way using the spectral synthesis technique with the FASMA\footnote{\url{https://github.com/MariaTsantaki/FASMA-synthesis}} software \citep{Tsantaki2018, Tsantaki2020}. The analysis with FASMA is based on the radiative transfer code, MOOG\footnote{\url{https://www.as.utexas.edu/~chris/moog.html}} \citep[version 2019,][]{sneden1973} and creates synthetic spectra on-the-fly to deliver the best-fit parameters after a non-linear least-squares fit (Levenberg-Marquardt algorithm). The line list is mainly comprised of iron lines initially obtained from VALD3 \citep[e.g.][]{Ryabchikova2015}. The atomic data ($\log gf$) were calibrated to match the spectrum of the Sun and Arcturus but with higher weights to the Sun which in this case could bias the results of our sample because it is comprised only of giant stars. Therefore, we used the suggested experimental $\log gf$ values from the recent line list of the {\em Gaia}-ESO survey \citep{Heiter2021}. The damping parameters are based on the ABO theory \citep{Barklem2000} when available, or in any other case, we use the Unsold approximation \citep{unsold1955}. The uncertainties are derived from the covariance matrix constructed by the nonlinear least-squares fit. 

In DM16, we delivered stellar parameters for 67 giants in 12 open clusters from HARPS and UVES spectrographs but with significant differences in the methods compared to this work. The methodology to obtain stellar parameters in our previous work was based on the ionisation and excitation balance of iron lines whose abundances were derived from their equivalent width measurements \citep[see also][]{tsantaki13, sousa2014}. In DM16, we used the Kurucz plane-parallel model atmospheres \citep{kurucz84} in local thermodynamic equilibrium (LTE). In this work, however, we use different model atmospheres, the MARCS models \citep{Gustafsson2008} which are better suited for the spherical atmospheric geometry of giant stars. Moreover, new spectra of several stars have been co-added from recent observing runs to increase their S/N. We set a low limit of 30 for the S/N to ensure reliability on the parameters with this method which mounts to 228 stars from the original sample. 

From the 43 targets with observations from both HARPS and UVES, we retrieved parameters from both instruments to examine any differences of our method with resolution. We find very good agreement for all parameters shown in Table~\ref{uves_harps} which allows us to include the ten stars with only spectra from UVES (see Fig.~\ref{param_comparison_spectrographs}) without jeopardising homogeneity. 

The comparison of the stellar parameters in this work with DM16 shows significant differences for the 67 stars in common, mostly in a form of an offset in all parameters, demonstrating the difficulty of the parameter derivation for giant atmospheres (see Fig.~\ref{param_comparison_dm16}). As we mentioned before, these discrepancies mostly arise from the different techniques and models atmospheres used. Because of these differences, we further compare our metallicities with literature works and we also include other methods to obtain T$_{\rm eff}$ and $\log g$ determinations in order to verify our results. 

We compare the metallicities with literature values obtained from other high resolution studies. First, we exclude non members from two different sources: non members from the WEBDA database\footnote{\url{https://webda.physics.muni.cz/webda.html}} and also by cross-matching with stars with low probability of belonging to the cluster (less than 60\%) from \cite{Cantat2020}. Then we calculate the mean metallicity of each cluster as shown in Table~\ref{feh_cluster_table} and plotted in Fig.~\ref{feh_cluster} along with literature values from high-resolution spectroscopic studies and for completeness we also add photometric metallicities for the rest of the clusters with no spectroscopic literature values. We compare our metallicities with the literature values taken only from high-resolution studies and find median difference of 0.05\,dex with standard deviation ($\sigma$) of 0.06\,dex and a median absolute deviation (MAD) of 0.03\,dex for the 28 OCs with literature values available, indicating very good agreement. 

Our spectroscopic temperatures were compared with T$_{\rm eff}$ derived from photometric calibrations. We use the relations between T$_{\rm eff}$ and colours including the effects of $[Fe/H]$ from \cite{Gonzalez2009}. We use the 2MASS magnitudes \textit{JHK$_{s}$} for the respective colour calibrations (V -- \textit{J}, V -- \textit{H}, V -- \textit{K$_{s}$}) from their equation 10 for giant stars which are the ones to exhibit the smallest standard deviations from the rest of their colour calibrations. Before applying the calibration, the colours were corrected for the reddening of the respective band, e.g. for V -- \textit{J}: 

\begin{equation}
(V - \textit{J})_{0} = (V - \textit{J}) - E(V - \textit{J}) = (V - \textit{J}) - A_{V} (1 - R) ,
\end{equation}
\\*
where A$_{v}$ is the extinction in the V magnitude calculated in Sect.~\ref{section_results} (Table~\ref{cluster_params}), and R is the relative extinction, R = $\frac{A_{J}}{A_{V}}$, taken from \cite{Wang2019}. The statistics from the comparisons for the three calibrations are described in Table~\ref{stellar_params_diff_table}. We have very small median differences in all filters (less than 38\,K) with the MAD reaching up to 106\,K for the (V -- \textit{J}) calibration. The results are also plotted in Fig.~\ref{teff_cluster} for the stars within the applicability limits of the calibrations. For the final photometric T$_{\rm eff}$ we use the average values from the three calibrations. 

More reliable surface gravities are obtained from methods with less model dependence than spectroscopy, such as asteroseismology or from dynamic mass and radius measurements in eclipsing binary systems. However, these measurements are limited to a relatively small number. With the {\em Gaia} mission \citep{Gaia2016}, we have access to parallaxes with unprecedented precision for millions of stars. In lack of other direct $\log g$ estimates, trigonometric gravities are very useful to evaluate our spectroscopic determinations. Trigonometric gravities are derived from the following expression: 

\begin{equation}\label{eq2}
\resizebox{.93\hsize}{!}{$ \log \frac{g}{g_{\odot}} = \log M + 0.4 (V + BC - A_V) + 4 \log \frac{T_{\rm photo, eff}}{T_{\rm eff \odot}} + 2 \log plx + 0.104 ~. $}
\end{equation}
 
The stellar masses (M) are derived from the PARAM 1.3 tool\footnote{\url{http://stev.oapd.inaf.it/cgi-bin/param\_1.3}} using the PARSEC theoretical isochrones \citep{bressan2012} and a Bayesian estimation method \citep{dasilva06}, the T$_{\rm eff}$ is the mean photometric temperature from the three colours derived above, and the metallicities are from the spectroscopic analysis of this work. We use the bolometric correction (BC) of \cite{Alonso1999} and V magnitudes from SIMBAD. Because of systematics in the {\em Gaia} DR2 parallaxes (plx) \citep{Gaia2018}, we add a conservative zero point value of 0.045\,mas proposed by the {\em Gaia} collaboration \citep{Lindegren2018}. Moreover, we increase the errors in parallaxes to consider the 30\% underestimation in uncertainties for bright stars \citep{luri2018, arenou2018}. The solar magnitudes are from \cite{bessell98}. The A$_{V}$ is calculated in Sect.~\ref{section_results} (Table~\ref{cluster_params}). The trigonometric gravities (for the stars with available masses) agree very well with our spectroscopic determinations showing --0.02\,dex median difference with very small $\sigma$ and MAD of 0.13 and 0.07\,dex respectively (see Fig.~\ref{logg_cluster} and Table~\ref{stellar_params_diff_table}). We present our spectroscopic parameters in Table~\ref{stellar_params_table_spec} as well as the parameters derived with the other methods described above in Table~\ref{stellar_params_table_photo}.
 
\section{Determination of Li abundances}\label{Li_determination}

\begin{table}
\centering
\caption{The Li line list.}
\label{li_linelist}
\begin{tabular}{lcc}
\hline\hline
Wavelength (\AA{}) & Excitation potential (eV) & Broadening\\
\hline
6707.920 &	0.00	& --0.807 \\								
6707.919 &	0.00	& --0.807 \\								
6707.908 &	0.00	& --0.807 \\								
6707.907 &	0.00	& --1.509 \\								
6707.768 &	0.00	& --0.206 \\								
6707.756 &	0.00	& --0.428 \\
\hline
\end{tabular}
\end{table} 

\begin{figure}
\centering
\includegraphics[width=1.0\linewidth]{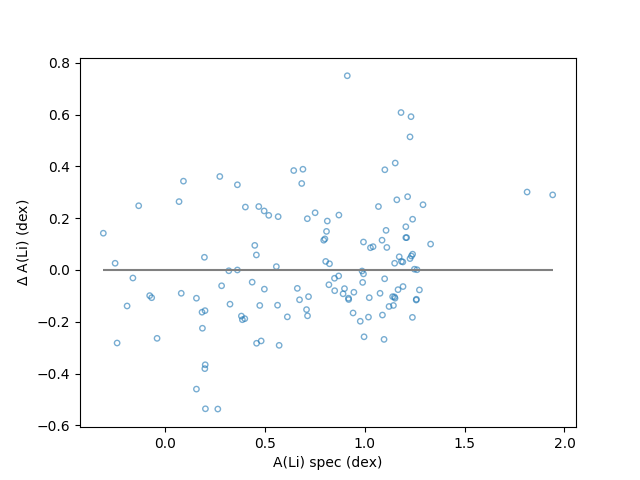}  \\ 
\caption{Comparison of Li abundances derived with spectroscopic parameters of this work and the ones with T$_{\rm eff}$ from photometry and trigonometric $\log g$.}
\label{li_spec_photo}
\end{figure}

\begin{figure}
  \centering
  \includegraphics[width=1.0\linewidth]{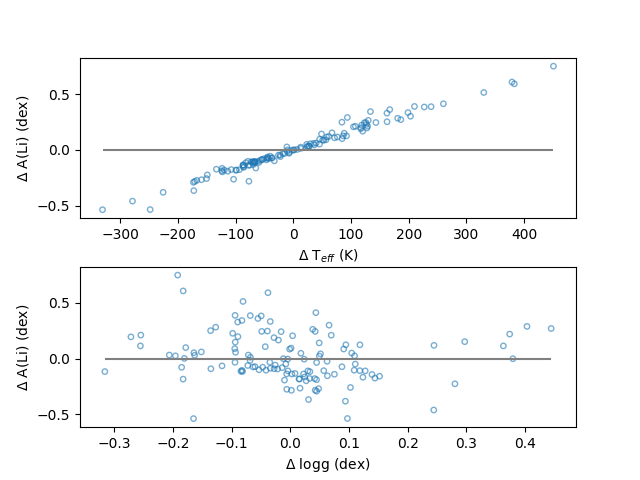}  \\ 
  \caption{Impact of stellar parameters on the determination of Li abundance. The y-axis in both plots is the difference in Li abundance derived from spectroscopic T$_{\rm eff}$ and $\log g$ and from photometric T$_{\rm photo, eff}$ and trigonometric $\log g$. Top plot: difference of Li abundance as a function of the difference in T$_{\rm eff}$ between spectroscopy and photometry. Bottom plot:  difference of Li abundance as a function of the difference in  T$_{\rm eff}$ between spectroscopy and trigonometric $\log g$. }
 \label{dli_dparams}
  \end{figure}

\begin{figure}
\centering
\includegraphics[width=1.0\linewidth]{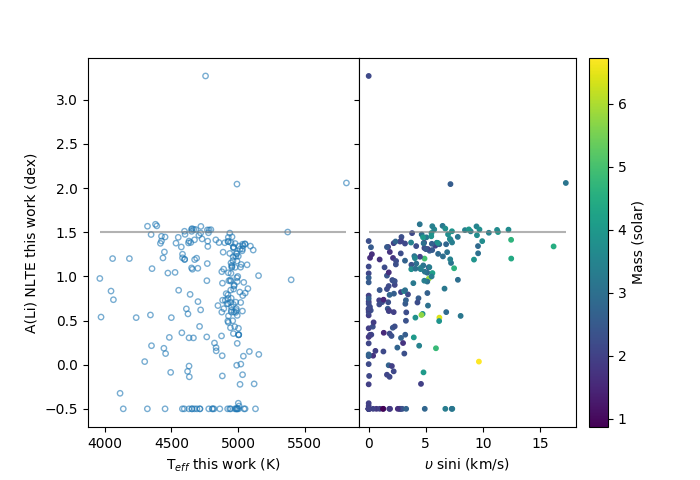}  \\
\caption{Li abundances as a function of T$_{\rm eff}$ (left panel) and rotational velocity colour coded to stellar mass (right panel). The horizontal line shows the canonical Li-rich limit.}
\label{li_nlte_teff_vsini}
\end{figure}

\begin{figure}
\centering
\includegraphics[width=1.0\linewidth]{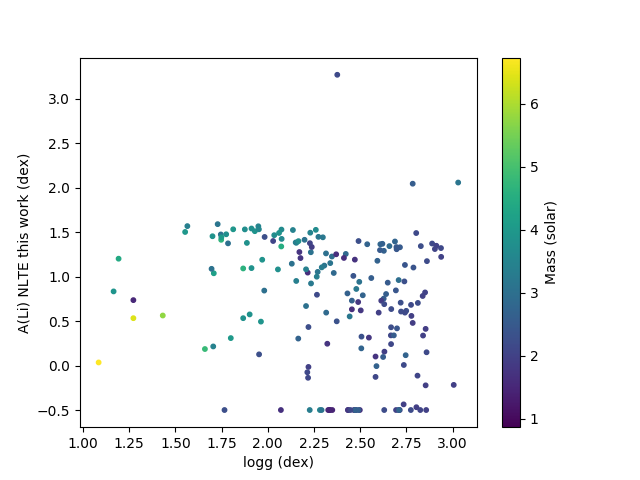}  \\
\caption{Li abundances as a function of $\log g$ and colour coded to stellar mass.}
\label{li_nlte_mass}
\end{figure}

The above results show that our stellar parameters are robust and we can further use their values for the determination of Li abundances. We derive LTE lithium abundances by performing spectral synthesis with FASMA \citep{Tsantaki2020} using the spectroscopic stellar parameters as fixed inputs and Li abundance as the only free parameter to solve the $\chi^{2}$ minimisation process. The line list is taken from \cite{Ghezzi2009} and is the same used in DM16, shown in Table~\ref{li_linelist}. The errors in Li abundance are obtained from the covariance matrix of the $\chi^{2}$ fit between the synthetic and observed spectrum. The most Li-poor stars reach the limit of our minimisation process (--0.5\,dex) and for them we provide this value as upper limit (noted in Table~\ref{stellar_params_table_spec}). The largest errors appear for the Li-poorer stars because their Li line is so weak that large changes in the abundance do not change the $\chi^{2}$ of the fit. 

Lithium abundances are known to be affected by non-LTE (NLTE) effects. \cite{Lind2009} have provided tabulated NLTE corrections for a grid of stellar parameters. After interpolating the grid of NLTE corrections, we find that all corrections are positive between 0.12\,dex and 0.38\,dex producing higher abundances, apart from the most Li-rich star (A(Li)\,=\,3.27\,dex) for which the correction is negative (--0.03\,dex). Apart from Li abundances derived from the spectroscopic values of our spectral analysis, we also calculate Li using a combination of spectroscopic parameters ($[Fe/H]$, microturbulence, macroturbulence, $\upsilon \sin i$), photometric T$_{\rm eff}$, and trigonometric $\log g$ from Sect.~\ref{method}. The overall differences in Li using stellar parameters from these two methods are in agreement with a median difference of --0.02\,dex and a standard deviation of 0.22\,dex (see Fig.~\ref{li_spec_photo}). The changes in T$_{\rm eff}$ and $\log g$ can affect the determination of Li abundances. In Fig.~\ref{dli_dparams}, we show the effect of using T$_{\rm eff}$ and $\log g$ from different methods on Li determinations. Lithium abundance is very sensitive to T$_{\rm eff}$ and significantly less to $\log g$. This sensitivity to stellar parameters is expected and is intrinsic to the spectroscopic methods either using the EW measurements or spectral synthesis to derive Li but for the other elements as well and showcases the biases which could come into play in the derivation of chemical abundances \citep[e.g.][]{Adibekyan2012}. Moreover, high rotation can introduce biases on the accurate determination of the A(Li) because the continuum placement becomes more difficult to place and severe line blending occurs. FASMA has been tested to provide reliable stellar parameters for stars up to $\sim$50\,km\,s$^{-1}$. Fortunately, the fastest rotator in this work reaches $\upsilon \sin i$\,=\,17.3\,km\,s$^{-1}$. In Fig.~\ref{high_vsini_fit}, we show the best-fit synthetic spectra and the observations around the Li lines for the three of the fastest rotators in our sample.   

Because it is difficult to claim which set of parameters (purely spectroscopic or spectro-photometric) is more accurate for the Li determinations and statistically both sets agree, we use the purely spectroscopic parameters derived in this work based on our experience in the derivation of stellar parameters with spectral synthesis \cite{Tsantaki2014, Tsantaki2018} to extract our further results. We also provide the Li abundances derived with the spectro-photometric input parameters in Table~\ref{stellar_params_table_photo}. Finally, we have 17 stars with Li enhanced values according to the canonical limit (A(Li)$>=$\,1.5\,dex) in NLTE using the spectroscopic parameters of this work. 

In the left panel of Fig.~\ref{li_nlte_teff_vsini}, we show the derived NLTE Li abundances as a function of spectroscopic T$_{\rm eff}$. We do not see any evident increase with T$_{\rm eff}$ for our temperature range ($\sim$4000-5200\,K) as opposed to the main sequence (MS) stars where Li abundance correlates with T$_{\rm eff}$ as a result of the reduced depletion of Li in the thinner convective envelopes of hotter stars \citep[e.g.][]{delgado15}. In the right panel of Fig.~\ref{li_nlte_teff_vsini}, we see an increasing trend of A(Li) with rotational period. A similar trend has been observed in other works \citep[][DM16]{Carlberg2012, DeMedeiros2000, Magrini2021b} where the authors find that "rapid" rotators ($\upsilon \sin i$ $>$\,8-10\,km\,s$^{-1}$) have on average higher lithium abundances when compared to the slow rotators. There is also a secondary dependence here on stellar mass, as lower-mass stars (and therefore older) have the lowest $\upsilon \sin i$ while the faster rotators are more massive (and younger) with higher Li abundances. We note though that our range in $\upsilon \sin i$ is not very wide to confirm this behaviour for faster rotators. According to \cite{Magrini2021b}, models which include rotation-induced mixing predict better the behaviour of Li in massive stars such as the ones in our sample, while for the lower mass stars, models based on thermohaline mixing can reproduce the observed Li. 

Finally, in Fig.~\ref{li_nlte_mass} we show the Li abundance as a function of $\log g$ and stellar mass. For stars with M\,$<$\,2.5-3.0M$_{\odot}$ there is a wide spread of Li abundances from Li poor to Li rich for stars at the same evolutionary stage. This behaviour for lower mass stars was also found by the GALAH survey \citep{Deepak2020}. For higher mass stars, the Li abundance is slightly higher indicating that mass could play a role in Li depletion. 

\section{Results}\label{section_results}

\begin{figure*}
 \centering
 \includegraphics[width=0.46\linewidth]{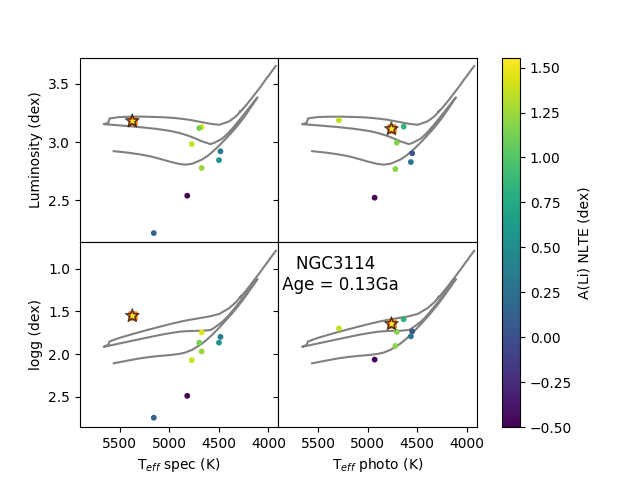} 
 \includegraphics[width=0.46\linewidth]{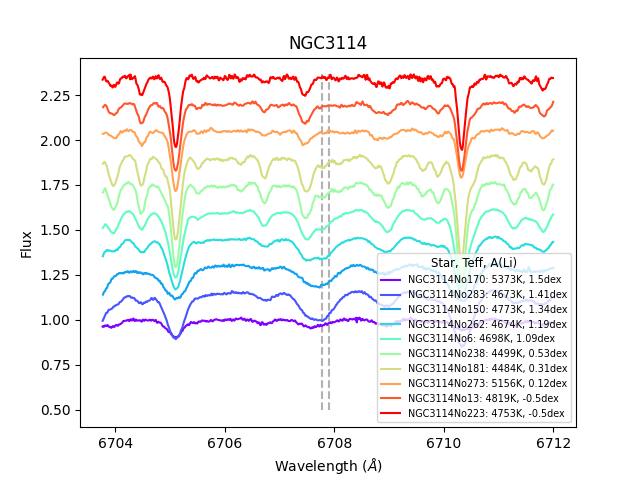} \\
 \includegraphics[width=0.46\linewidth]{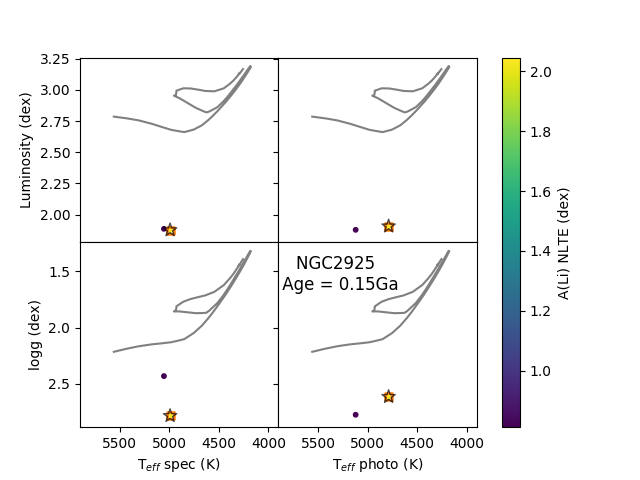}  
 \includegraphics[width=0.46\linewidth]{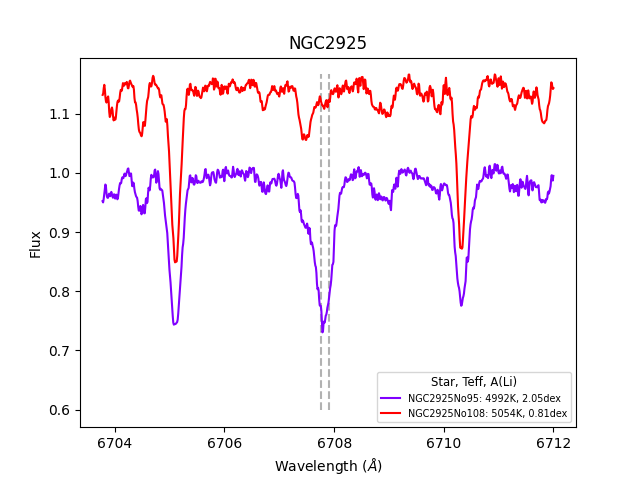}  \\ 
 \includegraphics[width=0.46\linewidth]{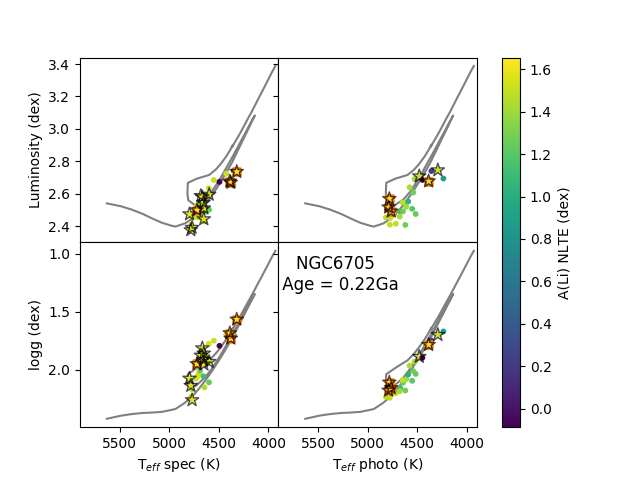} 
 \includegraphics[width=0.46\linewidth]{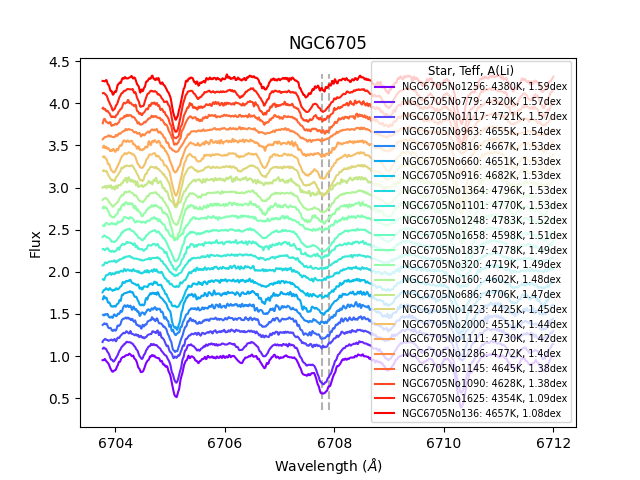}  \\
 \includegraphics[width=0.46\linewidth]{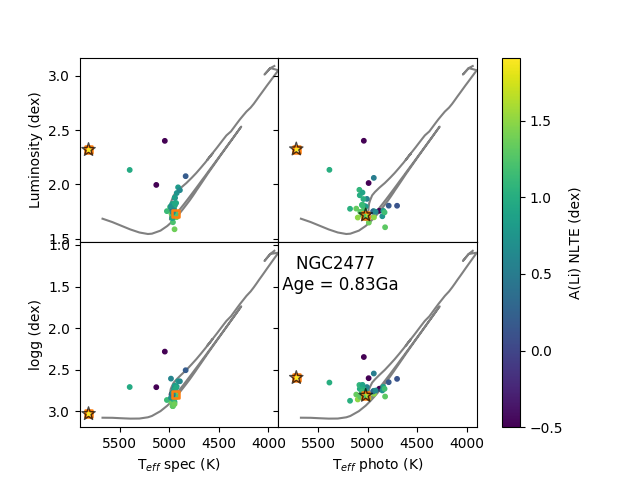}  ~~
 \includegraphics[width=0.46\linewidth]{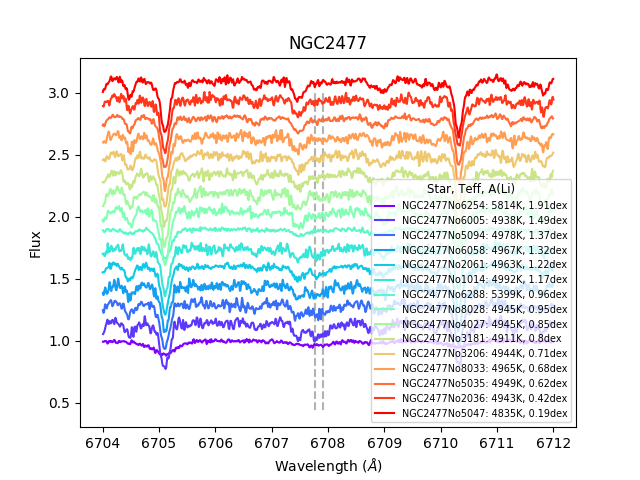}  
\caption{The HR diagrams of OCs with Li-rich members color coded to the Li abundance are plotted in the left panels with two sets of parameters: the spectroscopic parameters (left) and the photometric temperature and trigonometric $\log g$ (right). The isochrones are taken from the PARSEC tracks. The star symbols correspond to the canonical Li-rich stars. The orange square symbols correspond to the Li-rich stars which are defined with the criteria in Sect.~\ref{li_rich_clusters_b}. The right panels show the Li lines in their spectra. In the right panels of NGC6705 and NGC2477, we demonstrate the stars with the highest S/N for visual clarity.}
\label{hr_a}
\end{figure*}

\begin{figure*}
 \centering
 \includegraphics[width=0.46\linewidth]{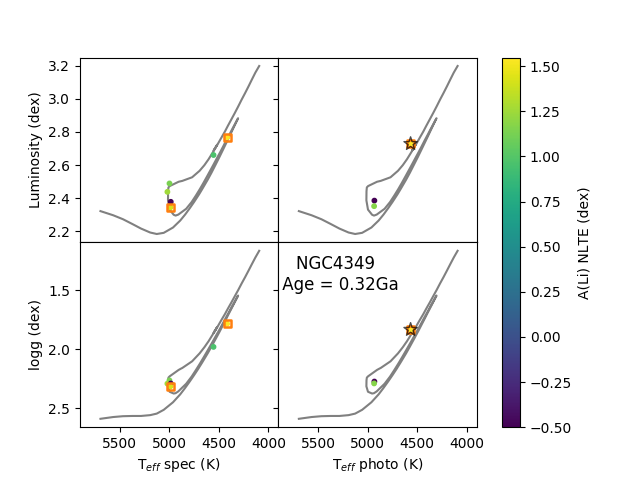}  
 \includegraphics[width=0.46\linewidth]{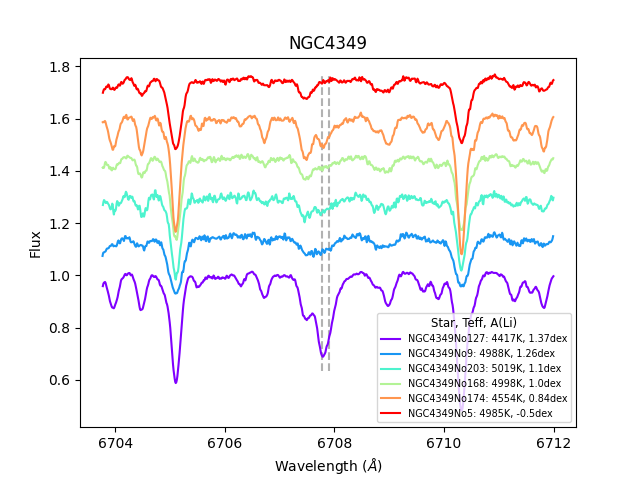}  \\
 \includegraphics[width=0.46\linewidth]{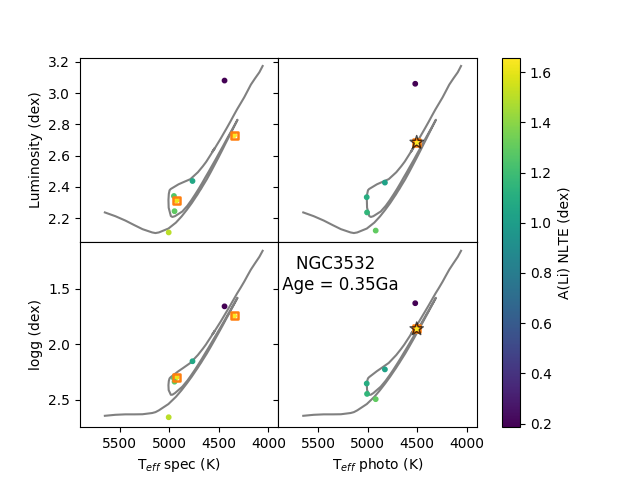}  
 \includegraphics[width=0.46\linewidth]{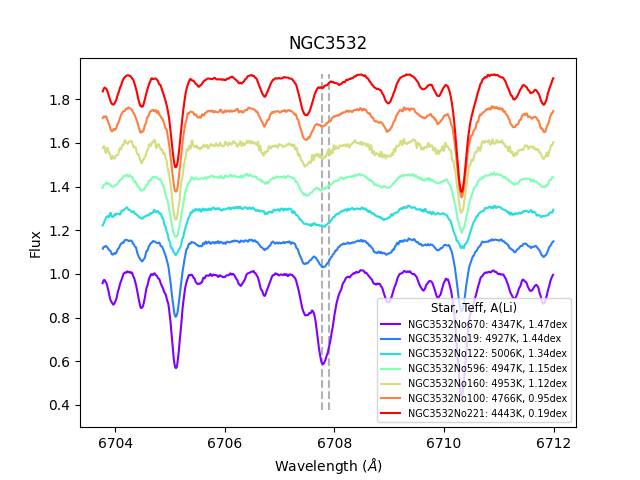}  \\
\includegraphics[width=0.46\linewidth]{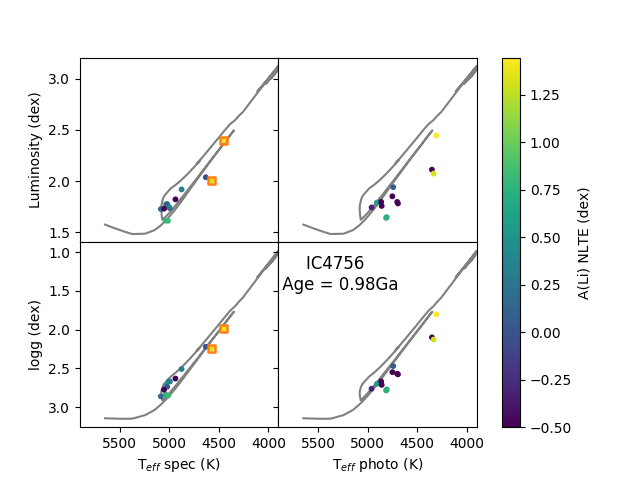} ~~
\includegraphics[width=0.46\linewidth]{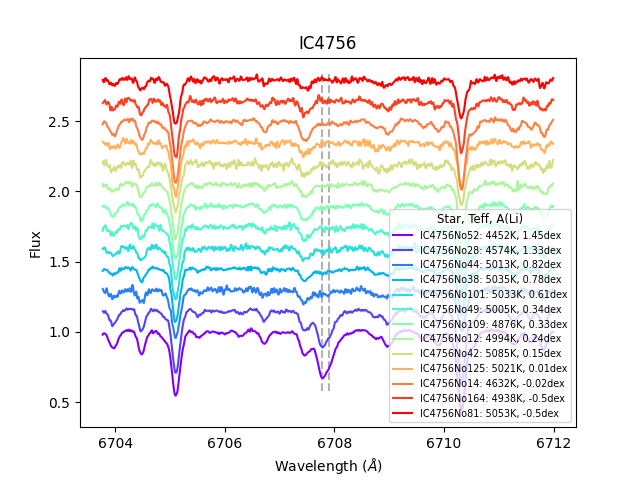}  
\caption{Same as Fig.~\ref{hr_a} but for the OCs with Li-rich giants defined with the criteria in Sect.~\ref{li_rich_clusters_b}.}
\label{hr_b}
\end{figure*}

\begin{table}
\centering
\caption{The distance modulus ((m--M)$_{0}$), extinction (A$_{V}$), ages, and the Li-rich limit estimation (A(Li)$_{max}$).}
\label{cluster_params}
\begin{tabular}{lcccc}
\hline\hline
Cluster & (m--M)$_{0}$ & A$_{V}$ & Age (Ga) & A(Li)$_{max}$ (dex)\\
\hline
IC2714  & 10.64 & 0.99 & 0.47 & 1.46 \\
IC4651  & 9.96  & 0.41 & 1.58 & 1.46 \\
IC4756  & 8.40  & 0.40 & 0.98 & 1.26 \\
NGC1662 & 7.84  & 0.62 & 0.91 & 1.46 \\
NGC2204 & 13.01 & 0.01 & 2.09 & 1.37 \\
NGC2251 & 10.71 & 0.72 & 0.31 & 1.46 \\
NGC2287 & 9.24  & 0.06 & 0.20 & 1.26 \\
NGC2345 & 12.32 & 1.91 & 0.07 & 1.37 \\
NGC2354 & 10.62 & 0.38 & 1.66 & 1.26 \\
NGC2360 & 10.24 & 0.11 & 1.12 & 1.26 \\
NGC2423 & 9.84  & 0.29 & 1.02 & 1.46 \\
NGC2447 & 10.09 & 0.11 & 0.56 & 1.26 \\
NGC2477 & 10.88 & 0.95 & 0.83 & 1.45 \\
NGC2506 & 12.54 & 0.17 & 2.14 & 1.37 \\
NGC2539 & 10.52 & 0.21 & 0.68 & 1.46 \\
NGC2567 & 11.35 & 0.47 & 0.34 & 1.46 \\
NGC2682 & 9.73  & 0.12 & 3.63 & 1.46 \\
NGC2925 & 9.38  & 0.29 & 0.15 & 1.46 \\
NGC2972 & 11.49 & 1.20 & 0.49 & 1.45 \\
NGC3114 & 10.11 & 0.26 & 0.13 & 1.46 \\
NGC3532 & 8.39  & 0.07 & 0.35 & 1.40$\dagger$ \\ 
NGC3680 & 10.15 & 0.20 & 1.78 & 1.26 \\
NGC4349 & 11.29 & 1.16 & 0.32 & 1.26 \\
NGC5822 & 9.72  & 0.35 & 0.89 & 1.46 \\
NGC6067 & 11.51 & 0.88 & 0.13 & 1.53$\dagger$ \\
NGC6134 & 10.36 & 0.87 & 1.45 & 1.45 \\
NGC6208 & 10.34 & 0.63 & 1.74 & 1.26 \\
NGC6281 & 8.63  & 0.42 & 0.32 & 1.46 \\
NGC6425 & 10.04 & 0.69 & 0.54 & 1.46 \\
NGC6494 & 9.35  & 0.85 & 0.38 & 1.46 \\
NGC6633 & 7.87  & 0.45 & 0.78 & 1.46 \\
NGC6705 & 11.84 & 1.10 & 0.22 & 1.54$\dagger$ \\
\hline
\end{tabular}
\begin{tablenotes}
\item The ages, distance modulus, and extinction are derived based on the methodology of \cite{Bossini2019} and the A(Li)$_{max}$ is an estimation on the maximum Li abundance estimated from \cite{Randich2020} and \cite{Romano2021}. $\dagger$ These values are taken directly from \cite{Randich2020} and \cite{Romano2021} (or their average) while the rest are estimations depending on the OC metallicity of this work.
\end{tablenotes}
\end{table} 

\subsection{Definition of a Li-rich giant}\label{definition_lirich}

As we mentioned before, Li enhancement is tied to the evolutionary stage of a star which can be considerably disentangled in the case of stars belonging to OCs. For the rest of the analysis, we exclude NGC2355 and NGC3960 which have only one member. We calculate the age of each cluster following the approach of \cite{Bossini2019}. The method is based on an automated Bayesian tool, BASE-9, to fit stellar isochrones on the observed G\,mag, and G$_{BP}$, G$_{RP}$ colours for the high-probability member stars. The resulting ages, distance modulus, and the extinction of the fits are in Table~\ref{cluster_params}. In Figs.~\ref{hr_a}-\ref{hr_b}, we show the HR diagrams and the $\log g$ - T$_{\rm eff}$ representation for the clusters which host Li-rich giants sorted by their age. The isochrones are from the PARSEC models \citep{bressan2012}\footnote{\url{http://stev.oapd.inaf.it/cgi-bin/cmd}} and the luminosities are derived using the distance modulus, the extinction, and the T$_{\rm eff}$ from spectroscopy of this work. 

The classification of a Li-rich giant in the literature usually is referred to a star with A(Li)\,$>=$\,1.5\,dex. This is considered the maximum canonical value and is derived assuming that the convective envelope of a solar metallicity giant star can induce a dilution by a factor of $\sim$60 and an abundance depletion of $\sim$1.8-1.9\,dex for a mass of 1.5\,M$_{\odot}$ assuming the initial interstellar medium abundance is $\sim$3.3\,dex \citep[e.g.][]{Charbonnel2000}. Therefore, the red giant may be expected to have a maximum surface abundance of 1.5\,dex for a 1.5\,M$_{\odot}$ star but this criterion in the case of OCs can be better constrained if we consider the Li of the progenitors of the Li-rich giants \citep[see discussion in][]{Sun2022}. We note that most of the giant stars in this sample have masses above 1.8-2.0\,M$_{\odot}$ and thus the determination of Li abundances for stars similar to their progenitors in the MS is not possible (the Li lines are not present in the spectra of so hot stars). However, these stars lay clearly at the hot side of the Li dip, regardless of their metallicity \citep[e.g.][]{delgado15} and thus no significant Li depletion is expected during the MS. Therefore, we can assume that their initial Li was the maximum possible in the cluster considering their age and metallicity. 

\cite{Randich2020} and \cite{Romano2021} provide maximum Li values for stars which have not undergone Li depletion yet in 18 and 26 OCs respectively based on homogeneous data from the {\em Gaia}-ESO survey \citep{gilmore2012, Randich2013}. Both works used the same criteria to select stars in each OC which have suffered minimal (or any) Li depletion and then derived their average maximum Li abundance. For instance, \cite{Randich2020} determine the maximum Li abundance of NGC6705 from members that have not suffered any Li depletion at 3.43\,dex which means that a Li-rich giant after Li dilution could be defined as a star having Li higher than $\sim$1.58\,dex (assuming a depletion of 1.85 as in \cite{Charbonnel2000}). \cite{Romano2021} report for the same cluster the maximum Li abundance 0.09\,dex less than \cite{Randich2020} which means that the derived limits of Li-rich giants can fluctuate between 1.49-1.58\,dex. If we consider the average value of both works, from the 13 canonical Li-rich giants we have found in NGC6705, only five are above the limit to be considered as Li rich (see Sect.~\ref{li_rich_clusters_a}) but this number can vary significantly between one (with the 1.58\,dex limit) and 15 (with the 1.49\,dex limit).

Due to the lack of measurements of these limits in the literature (only one OC in common with \cite{Randich2020} and three with \cite{Romano2021}), we use both data sets to make a rough extrapolation on the maximum Li abundance for our clusters. We divide the maximum Li abundance of the above literature samples into metallicity bins of 0.1\,dex in size (see Fig~\ref{limax}). We use our iron metallicities of our OCs to obtain the maximum Li abundance which is assigned as the average of the two studies for the respective metallicity bin. The results are presented in Table~\ref{cluster_params} and are an approximation on the limits of the Li-rich giants based on the evolution of their progenitors for these clusters \cite[see also ][]{Sun2022}. 

We examine below case by case the clusters which own Li-rich giants set by the canonical limit (Sect.~\ref{li_rich_clusters_a}), and on the criterion based on the Li evolution of their progenitors mentioned above (Sect.~\ref{li_rich_clusters_b}) using the Li abundances derived from our purely spectroscopic parameters. Moreover, we show special cases of giants with strong Li lines in their spectra (Sect.~\ref{li_rich_clusters_c}) sorted by age.

\subsection{Clusters with canonical Li-rich giants}\label{li_rich_clusters_a}

\subsubsection{NGC3114}

This young cluster has one Li-rich star, NGC3114No170, with Li abundance of 1.50\,dex. This star is quite massive (4.0\,M$_{\odot}$) and has passed the RGB phase. It is located at the red clump where He is burning in its core and has not reached the AGB phase yet which is an evolutionary stage to produce Li via the HBB process. It exhibits relatively high $\upsilon$ $\sin i$ (11\,km\,s$^{-1}$) and in fact, in this cluster the Li abundance correlates well with $\upsilon$ $\sin i$ which is also evident from the broadening of the spectral lines in Fig.~\ref{hr_a}. For instance, the three fastest rotators in this cluster (NGC3114No150, NGC3114No283 and NGC3114No170) are the ones with the highest Li abundance in this cluster (A(Li)\,=\,1.34, 1.41, 1.51\,dex, respectively). Such high velocities in giants are not common because they slow down as they evolve away from the MS. 

We have noticed though that the metallicity of NGC3114No170 (--0.66\,dex) is far from the cluster average (--0.10\,dex). This indicates that the star may not be a member even though the kinematic criteria do not exhibit this. The same conclusion was also proposed by \cite{Santrich2013}, suggesting that this star belongs to the thick disc/Galactic halo because the metallicity and the abundances are very different from the other giants in this cluster. 

The high rotational velocities found in the stars above with the highest Li abundances in this cluster could be indicators of planet accretion which can explain both Li enhancement and rapid rotation as proposed in several works \citep[e.g.][]{Carlberg2012, Aguilera2016, Soares2021} even though these process is more likely to happen at the turnoff and on the lower giant branch. On the other hand, \cite{Casey2019} propose that stars in the core-helium-burning phase as in our case here can enhance Li if they exist in a tidal locked binary system. There is no evidence however that any of these stars are in binary systems. 

\subsubsection{NGC2925}

The Li-rich star in this cluster is NGC2925No95 (HD\,298542) with very high Li abundance of 2.06\,dex. However, it falls very far from the cluster isochrone and it is listed as binary star by The Washington Visual Double Star Catalog \citep{Mason2001}. Tidal interactions in binary systems can be strong enough to drive internal mixing triggering the CF mechanism which in turn can produce the observed high lithium abundance in this star \citep[e.g.][]{Casey2019}. Nevertheless, the work by \citet{Mermilliod2008} does not list this star as a binary (based on a few RV measurements taken during a span of $\sim$\,6 years) and the RVs from HARPS only show 60\,m\,s$^{-1}$ variability within a time span of 600\,days. Thus, more data would be needed to confirm a physical companion to this star. 

The incapability to determine the evolutionary stage of this star prevents us to know whether it has passed the FDU or not which could explain its high Li. However, the proximity in stellar parameters with the other cluster member with clear depleted Li (No108) makes us suspect that this star has suffered a special event altering its Li abundance.

\subsubsection{NGC6705} 

This cluster is young and contains 13 stars with Li abundances above the Li-rich canonical limit (between 1.50-1.59\,dex): No669, No1658, No1248, No1101, No1364, No916, No660, No816, No963, No1117, No779, No411, and No1256. In our sample, 27/29 have A(Li)$>$1.0\,dex and only two are Li poor. This may indicate that the environment of their formation is already abundant in Li and their MS progenitors could have had already high Li before diluted to their current values. As we mentioned in Sect.\ref{definition_lirich}, if we consider the maximum Li abundance calculated in Table~\ref{cluster_params}, then the limit of a Li-rich giant in this cluster could be defined $\sim$\,1.54\,dex. In this case, from the 13 canonical Li-rich giants in our sample, only five (No1256, No411, No779, No1117, and No963) are above the 1.54\,dex limit. This cluster despite being young is also enhanced in $\alpha$ elements \citep{Magrini2014, Casamiquela2018} and the existence of $\alpha$-rich stars at such a young age is not predicted by chemical evolution models. This is interesting evidence to showcase the peculiar birth environment of this cluster. 

The masses of these stars are above $\sim$3.0\,M$_{\odot}$. The HR diagram does not show clearly the evolutionary stage of the Li-rich giants. In the plot $\log g$ - T$_{\rm eff}$ with spectroscopic parameters, most of them (11/13) appear at the red clump and the rest (2/13) they start their first ascent at the RGB phase in Fig.~\ref{hr_a}. However, in the luminosity - T$_{\rm eff}$ plot, they all appear ascending the RGB. The Li-poorer stars in this cluster have lower rotational velocities compared to the more Li abundant. 

\subsubsection{NGC2477}

In this cluster, NGC2477No6254 (CD-38\,3745) has Li abundance of 1.91\,dex. The spectroscopic parameters show a G-type sub-giant star with relatively high rotation ($\upsilon$ $\sin i$\,=\,17.3\,km\,s$^{-1}$). This star has not experienced yet the FDU and its high Li abundance is thus expected. However, this star is interestingly classified as a yellow straggler \citep{Rain2021}. Yellow straggler stars are usually located between the main sequence and the giant branch (Hertzprung gap). They are identified as rather evolved blue stragglers.

In the same cluster, NGC2477No6005 exhibits also high Li abundance 1.49\,dex and is above the limit if we consider maximum Li abundance from Table~\ref{cluster_params} which is 1.45\,dex for this cluster. This star has a mass of 2.1\,M$_{\odot}$ and appears to be at the red clump phase (Fig.~\ref{hr_b}). This star has passed the FDU and probably has experienced the He flash so a mechanism for Li enrichment is needed. This is in agreement with observational data from large spectroscopic surveys which show that low mass (less than 2\,M$_{\odot}$) Li-rich giants are mostly found at the RC phase \citep[e.g.][]{Kumar2020, Yan2021, Martell2021}. The exact physical mechanism to cause enrichment at this stage is still not defined and probably is related to the He flash at the RGB tip before the star ends at the RC but the proposed models for including additional mixing processes have only been tested to explain the behaviour of Li in low-mass stars ($\sim$\,1\,M$_{\odot}$). 

\subsection{Clusters with Li-rich giants based on the Li abundance of their progenitors}\label{li_rich_clusters_b}

\subsubsection{NGC4349}

The Li-rich star of this cluster is NGC4349No127 with Li abundance 1.37\,dex (or 1.55\,dex if we consider the spectro-photometric parameters) and is well above the Li limit of 1.26\,dex for this cluster from Table~\ref{cluster_params}. This star shows RV variations corresponding to a signal of a sub-stellar companion \citep{Lovis2007}. However, more recent studies have showed that this signal can be mimicked by stellar activity modulation \citep{Delgado2018}. The origin of the RV variations has not yet been concluded and new data is required to comprehend if there is a companion or not. 

Nevertheless, this star is post-FDU located at the tip of the RGB phase or even on the way for the second ascend (Fig.~\ref{hr_b}). This star shows a clear strong Li line and we also classify it as Li rich, in the same way as was done in DM16, by comparison of their Li spectral lines with stars of similar parameters. In particular, NGC4349No127 (T$_{\rm eff}$\,=\,4417\,K, $\log g$\,=\,1.79\,dex, $[Fe/H]$\,=\,--0.17\,dex) has similar stellar parameters and age with NGC3532No670 (T$_{\rm eff}$\,=\,4347\,K, $\log g$\,=\,1.75\,dex, $[Fe/H]$\,=\,--0.11\,dex) and NGC3532No221 (T$_{\rm eff}$\,=\,4443\,K, $\log g$\,=\,1.66\,dex, $[Fe/H]$\,=\,--0.12\,dex). Both NGC4349No127 and NGC3532No670 (we also classify it as Li rich) have very strong Li lines in comparison to NGC3532No221 which is Li poor. Comparing the Li lines of stars with the same stellar parameters and at the same evolutionary stage is an evident criterion for classification of Li richness.

The mass of this star (3.0\,M$_{\odot}$) suggests it is too massive to experience LB but it is not yet at the AGB phase to experience HBB, a mechanism to trigger the CF mechanism. If this star is at the RGB tip, it is difficult to explain its Li enhancement due to intrinsic mechanisms. The possible presence of a companion could cause tidal interactions to spin up the primary star and increase the surface Li abundance at the RGB phase \citep{Casey2016}. Assuming this star is ascending the early-AGB phase, extra-mixing mechanisms can be evoked to cause Li enhancement \cite{Charbonnel2000}. 

In the same cluster, NGC4349No9 is also an interesting case as its Li abundance falls exactly on the limit to be considered Li rich (A(Li)$_{max}$\,=\,1.26\,dex) based on Table~\ref{cluster_params}. It has the highest rotational velocity from the six stars in this cluster ($\upsilon$ $\sin i$\,=\,9.6\,km\,s$^{-1}$) which is also demonstrated by the broadened spectral lines in Fig.~\ref{hr_b}. It is located in the RC but due to its mass (3.0\,M$_{\odot}$) it will not pass the LB and thus the exact physical mechanism to cause enrichment at this stage is still not defined. The Li enhancement could be related to rotation though to an external mechanism such as planet accretion or binarity.

\subsubsection{NGC3532} 

This cluster has one canonically Li-rich star, NGC3532No649 (HD\,95799) with Li abundance 3.27\,dex. This star is a known super Li-rich giant \citep{Luck1994} which also shows a low value of $^{12}$C/$^{13}$C, suggesting that it has suffered mixing and its Li has been freshly synthesized. However, we note that it is not classified as a cluster member after a kinematic selection \citep{Mermilliod2008} and thus it is not plotted in Fig.~\ref{hr_b}.

If we consider the maximum Li abundance of the progenitors to be 1.40\,dex (see Table~\ref{cluster_params}), then according to this criterion, NGC3532No670 is classified as Li-rich giant (A(Li)\,=\,1.47\,dex). The location of NGC3532No670 in the HR diagram, depends on which set of parameters we choose to interpret the results. If we consider purely spectroscopic parameters (left panel of Fig.~\ref{hr_b}), the star appears to have passed the first ascent and be close to the RGB tip and thus, Li depletion should be maximum. Because this star is massive enough (3.0\,M$_{\odot}$), it will not experience He-flash nor LB which are the evolutionary stages possible for triggering Li enhancement proposed in the literature for low-mass stars. The photometric T$_{\rm eff}$ is higher than the spectroscopic which places the star on the second ascend at the early-AGB phase (right panel of Fig.~\ref{hr_b}). At this phase the convective envelope deepens and an extra-mixing process may be most effective in bridging the small gap between the base of the convective zone and the H burning shell \cite{Charbonnel2000}. Then Li production is possible via the CF mechanism, explaining its high Li abundance. 

\cite{Smiljanic2009} have analysed the same cluster and classified this star as an early AGB based on high-resolution spectra. Moreover, they find a relatively high $^{12}$C/$^{13}$C ratio for NGC3532No670 implying that this star has not suffer an extra-mixing episode. This raises a difficulty in explaining the high Li abundance we observe in this case. 
We also note that the stellar parameters and age of this star are quite similar to NGC4349No127 which is also a Li-rich giant in this context and occupy the same positions in the HR diagrams.

\subsubsection{IC4756}

The cluster IC4756 has two stars (IC4756No52 and IC4756No28) with Li abundances (1.45 and 1.33\,dex respectively) which differs significantly compared to the rest of the stars with similar parameters, e.g. IC4756No14. These stars are Li-rich-in-context and lie on the RGB, with IC4756No52 closer to the RGB tip (Fig.~\ref{hr_b}). The Li-rich limit, based on metallicity, for giant stars in this cluster is placed at $\sim$1.26\,dex which justifies our classification of these stars as Li-rich giants. We note that \citet{Smiljanic2009} measured a low carbon isotopic ratio for IC4756No28, indicating that this star has probably undergone internal mixing which in turn could have supported the CF mechanism.

\subsection{Clusters with giants with strong Li lines}\label{li_rich_clusters_c}

Apart from identifying Li-rich giants using the criteria above, there are giant stars which exhibit high Li abundance compared to the rest of the stars of the same evolutionary stage in the same cluster. In fact, we saw in the previous Section cases where giant stars have Li abundances lower than the canonical limit but abnormally high A(Li) compared to other relevant stars. These stars can be defined as "Li-rich-in-context", definition taken from \cite{Sun2022}, and below we comment on such cases where giant stars show strong Li lines but they are below the two aforementioned limits. 

\begin{figure*}
\centering
\includegraphics[width=0.46\linewidth]{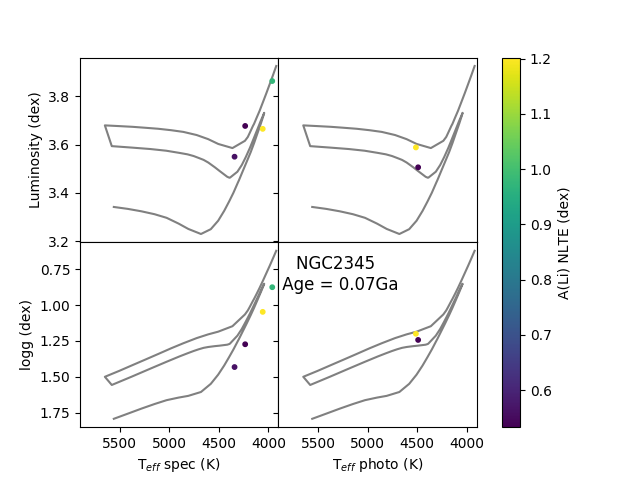} ~~
\includegraphics[width=0.46\linewidth]{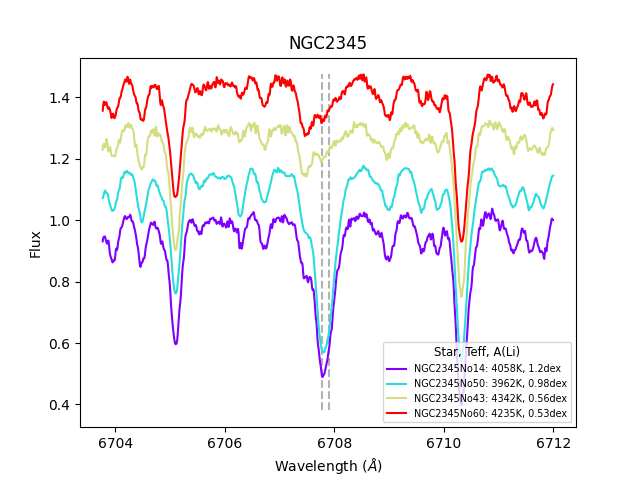}  
\includegraphics[width=0.46\linewidth]{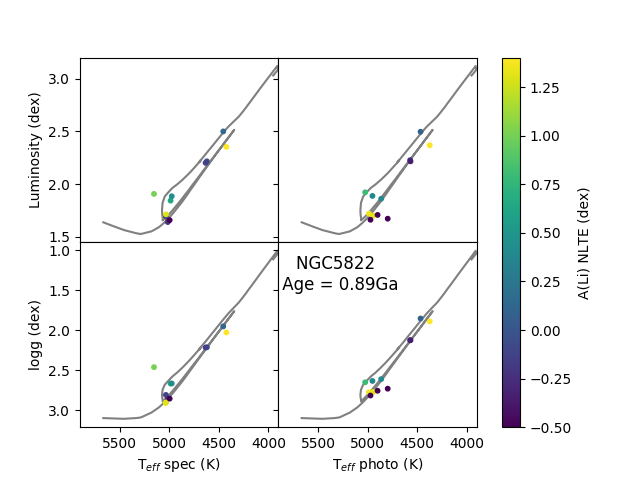} ~~
\includegraphics[width=0.46\linewidth]{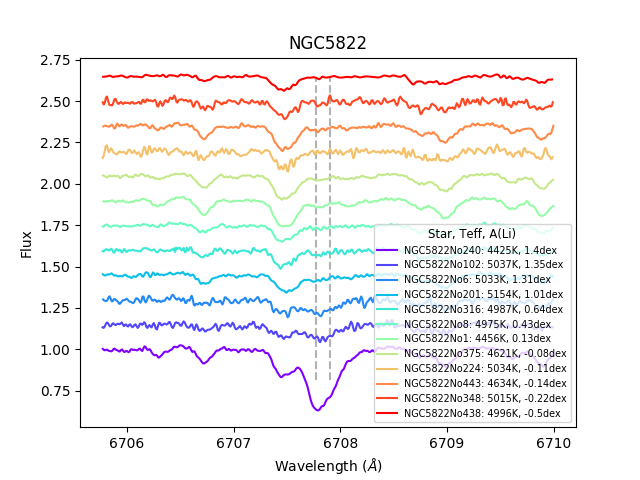}  
\includegraphics[width=0.46\linewidth]{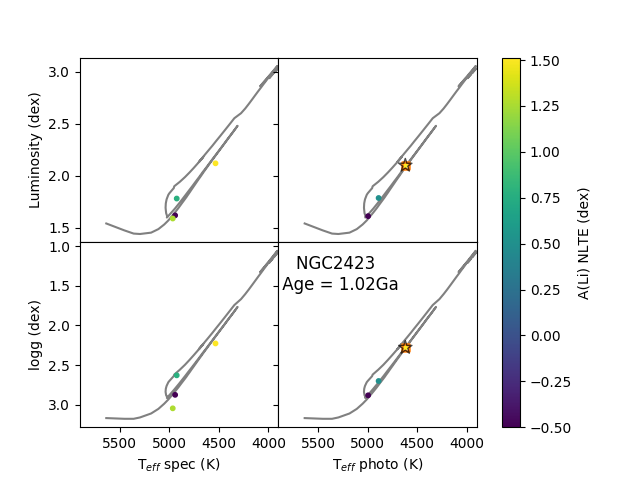}  
\includegraphics[width=0.46\linewidth]{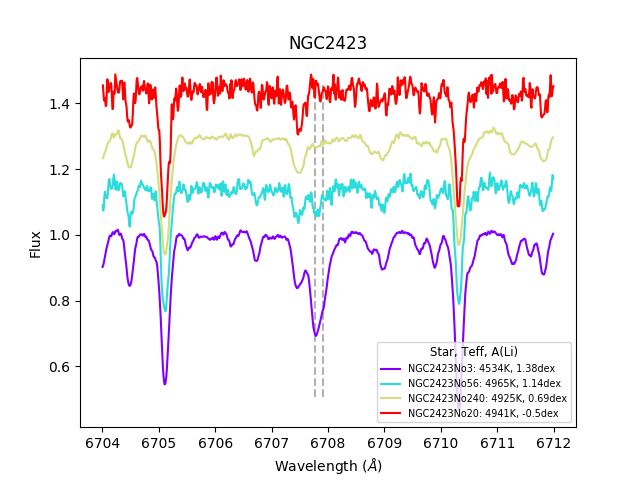} \\  
\includegraphics[width=0.46\linewidth]{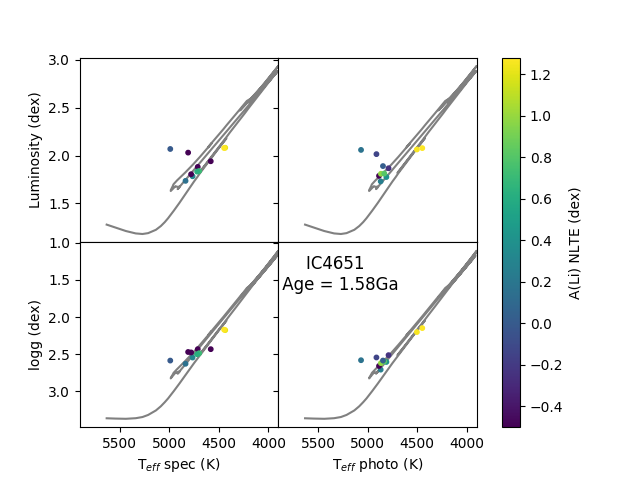} ~~
\includegraphics[width=0.46\linewidth]{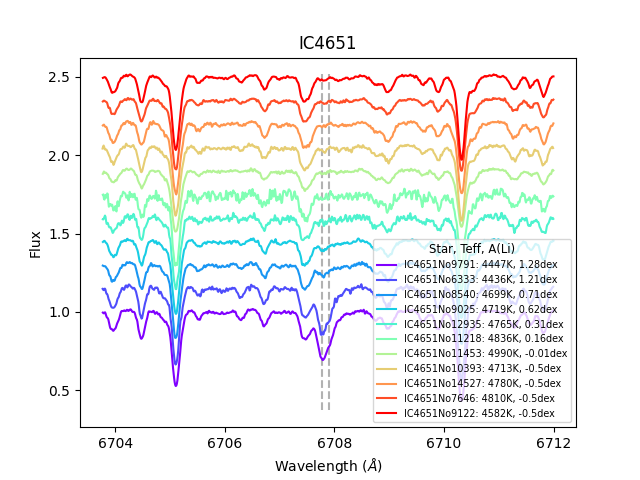}  
\caption{Same as Fig.~\ref{hr_a} for the giant stars of Sect.~\ref{li_rich_clusters_c}.}
\label{hr_c}
\end{figure*}

\begin{figure*}
\centering
\includegraphics[width=0.46\linewidth]{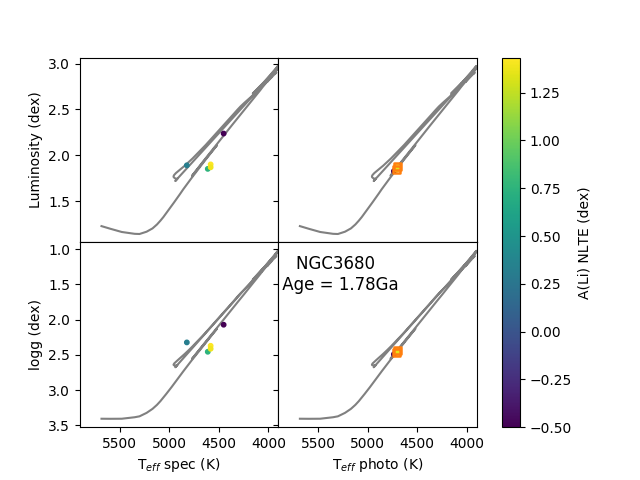} ~~
\includegraphics[width=0.46\linewidth]{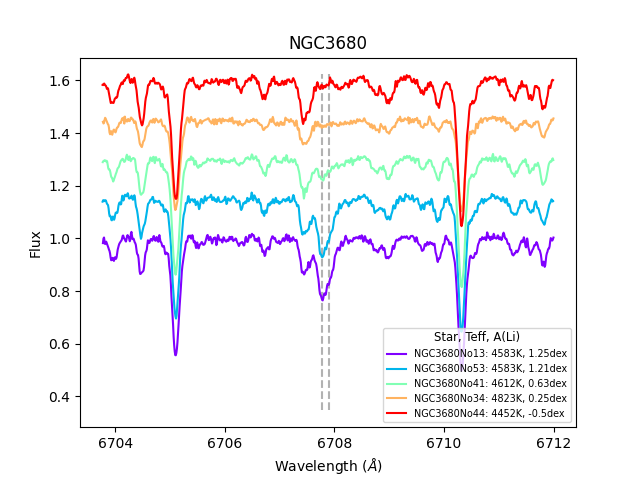}  
\caption{Same as Fig.~\ref{hr_a} for the giant stars of Sect.~\ref{li_rich_clusters_c}.}
\label{hr_d}
\end{figure*}

\subsubsection{NGC2345}

This is the youngest in our sample hosting four stars in our sample which have masses $\sim$\,5-6\,M$_{\odot}$. Two of these stars (NGC2345No14 and NGC2345No50) show high Li content compared to the other two at the same evolutionary stage. The Li abundances of NGC2345No14 and NGC2345No50 are 1.20 and 0.98\,dex respectively. 

From Fig.~\ref{hr_c}, these stars occupy the top part of the RGB phase based on spectroscopic parameters (left panel). According to \cite{Alonso2019}, their photometric analysis places these stars also at the RGB evolutionary phase in accordance to our spectroscopic parameters. Moreover, they measure A(Li) above the canonical limit for NGC2345No14 and NGC2345No50 but the work of \cite{Holanda2019} shows better agreement in our Li determinations for these stars. From their position in the HR diagram, there is no clear internal physical mechanism to favour some stars with enrichment and others not. On the other hand, external mechanisms such as the engulfment of a planet or a brown dwarf by these two stars could be a plausible explanation, but our data cannot confirm this or not. We do not notice higher rotational velocities which could also support this scenario. 

\subsubsection{NGC5822}

There are three stars in NGC5822 with high Li content: NGC5822No102, NGC5822No240, and NGC5822No6 (1.34, 1.40, and 1.31\,dex respectively). Because of the low iron metallicity of NGC5822, the Li-rich limit could fall around 1.46\,dex which is closer to the Li abundance of our three stars compared to the canonical limit. From Fig.~\ref{hr_c}, the most Li-rich star, NGC5822No240, is located at the RGB tip and its Li lines are evidently stronger compared to stars with similar parameters such as NGC5822No1. The other two giants could either appear at the base of the giant branch which could indicate that they have not finished the FDU or they maybe be already located in the clump. The work of \cite{Smiljanic2009} provides a carbon isotopic ratio of 17 for the NGC5822No240 (and no analysis for the other two) which may indicate some mixing and thus explain the higher Li.

\subsubsection{NGC2423}

We have analysed four stars in this cluster with masses $\sim$2\,M$_{\odot}$, meaning they may undergo the luminosity bump. The Li-rich star here, NGC2423No3 (BD-13\,2130), is also presented in DM16 and has Li abundance of 1.38\,dex using purely spectroscopy parameters or 1.51\,dex from spectro-photometric parameters (both in NLTE). This star is still below the limit of 1.46\,dex based on metallicity from Table~\ref{cluster_params} but it has a clear high A(Li) compared to the other stars in the cluster. The evolutionary stage of this star is in fact close to the LB which might allow the extra-mixing to trigger the CF mechanism at this phase. The isochrone of NGC2423 does not show this stage probably because the derived age is slightly younger and corresponds to a higher mass, a bit over the limit to experience the LB for this cluster (Fig.~\ref{hr_c}).

Interestingly, the same star might be the host of a giant planet (M$_{p}$=10.6\,M$_{J}$) at a distance of 2.1\,AU and with a period of 714\,days \citep{Lovis2007} although the planetary nature of the RV signals have been disputed in \citep{Delgado2018}. Under the scenario of Li enhancement as a consequence of planetary material accretion, in DM16 we speculate with the possibility that an hypothetical short-period planet in the system which could have existed given that there was already a probe of planet formation around that star might have been engulfed and hence produced the Li enhancement. However, this scenario was found to be not very probable considering the mass and evolutionary stage of the star.

\subsubsection{IC4651}

The cluster IC4651 has two stars with high Li abundance. Even if these stars are not canonically Li rich, IC4651No9791 and IC4651No6333 show strong Li lines (Fig.~\ref{hr_c}) with abundances of 1.28 and 1.21\,dex. Both stars fall on the LB and have a smaller mass compared to the rest of this cluster. As also noted in DM16, both stars may be in a Li enrichment process without having reached the highest Li abundance yet, or they may already be starting to destroy the Li created during the luminosity bump. If we consider the Li-rich limit we calculated for this cluster (1.46\,dex), both stars are still below it.

We also note that a star of this cluster (IC4651No9122) is the host to a planet candidate with a minimum mass of 7.2\,M$_{J}$ and a period of 747 days \citep{Delgado2018}, awaiting for further data for confirmation. This star is a Li-poor giant. 

\subsubsection{NGC3680}

NGC3680 has two stars (NGC3680No13 and NGC3680No53) with high Li abundance (1.25 and 1.21\,dex) compared to the rest of the cluster. The LTE abundance of NGC3680No13 is in perfect agreement with the work of \cite{AnthonyTwarog2009} but NGC3680No53 is underestimated by 0.18\,dex in our work. Both of these stars fall at the LB as in IC4651 which offer the same possible scenario (Fig.~\ref{hr_d}). However, there is a third star which also falls at the LB and has only half the abundance of the other two, meaning that it may be just starting the enrichment process at this evolutionary stage. The Li-rich limit we provide for this cluster (1.26\,dex) is close to the observational criteria we find for the classification of these giants as Li rich.

\section{Discussion and conclusions}\label{conclusions}

\begin{table*}
\centering
\caption{Classification of Li-rich giants derived in this work based on two criteria: the canonical limit and the maximum A(Li) defined by the progenitors of each OC ($>$ A(Li)$_{max}$ limit).}
\label{lirich_params}
\begin{tabular}{lcccc}
\hline\hline
Star & A(Li)$_{NLTE}$ (dex) & canonical limit & $>$ A(Li)$_{max}$ limit & Evolutionary stage \\
\hline
NGC3114No170$\dagger$ & 1.50 $\pm$ 0.03	& yes	& yes	& - \\
\hline
NGC2925No95 & 2.06 $\pm$ 0.01	& yes	& yes	& inconclusive \\
\hline
NGC6705No1101 & 1.53 $\pm$ 0.02	& yes	& no  & RGB \\
NGC6705No1117 & 1.57 $\pm$ 0.02	& yes	& yes & RC \\
NGC6705No1248 & 1.52 $\pm$ 0.03	& yes	& no  & RC \\
NGC6705No1256 & 1.59 $\pm$ 0.02	& yes	& yes & RGB \\
NGC6705No1364 & 1.53 $\pm$ 0.02	& yes	& no  & RC \\
NGC6705No1658 & 1.51 $\pm$ 0.02	& yes	& no  & RC \\
NGC6705No411 & 1.57 $\pm$ 0.02	& yes	& yes & RC \\
NGC6705No660 & 1.53 $\pm$ 0.03	& yes	& no  & RC \\
NGC6705No669 & 1.51 $\pm$ 0.03	& yes	& no  & RC \\
NGC6705No779 & 1.57 $\pm$ 0.02	& yes	& yes & RC \\
NGC6705No816 & 1.53 $\pm$ 0.03	& yes	& no  & RC \\
NGC6705No916 & 1.53 $\pm$ 0.03	& yes	& no  & RC \\
NGC6705No963 & 1.54 $\pm$ 0.03	& yes	& no  & RC \\
\hline
NGC2477No6005 & 1.49 $\pm$ 0.03	& no	& yes & RC \\
NGC2477No6254$\ddagger$ & 1.91 $\pm$ 0.03	& yes	& yes & RGB base \\
\hline
NGC4349No127 & 1.37 $\pm$ 0.02	& no	& yes	& RGB tip \\
NGC4349No9   & 1.26 $\pm$ 0.03	& no	& yes	& RC \\ 
\hline
NGC3532No649$\dagger$ & 3.27 $\pm$ 0.02	& yes	& yes & - \\
NGC3532No670 & 1.47 $\pm$ 0.02	& no	& yes	& RGB tip \\
NGC3532No19	 & 1.44 $\pm$ 0.02	& no	& yes	& inconclusive \\
\hline
IC4756No28	& 1.33 $\pm$ 0.02	& no	& yes	& RGB tip \\
IC4756No52	& 1.45 $\pm$ 0.02	& no	& yes  & RGB tip \\
\hline
\% of the OCs sample    & & 14/215=6.5\% & 12/215=5.6\%  & \\
\% of the total sample  & & 16/228=7.0\% & 14/228=6.1\%  & \\
\hline
\end{tabular}
\tablefoot{$\dagger$Star in defined as non member. $\ddagger$Star is located at the RGB base and it is not considered as Li rich.}
\end{table*} 

In this work, we present a sample of 247 stars belonging to 32 open clusters of which we provide detailed spectroscopic analysis of 228 stars, with 0.07\,Ga\,$<$\,ages\,$<$\,3.6\,Ga spanning a wide range of masses between 1 and 6\,M$_{\odot}$. We used the spectral synthesis technique to derive their Li abundances from their high resolution spectra. We have identified 14 giants to be Li enriched based on the classical A(Li) limit ($>$\,1.50\,dex) which have passed the FDU phase. We have provided estimations for the Li limit of enrichment for giants considering the maximum Li values these stars would have before depleting Li at the FDU, based on the works of \cite{Randich2020} and \cite{Romano2021}. Using the latter criterion, we find 12 Li-rich stars distributed amongst six different OCs based on our spectroscopic stellar parameters. 

In total, the percentage of Li-rich giants corresponds to $\sim$6-7\% of our whole sample depending on the criteria for the classification and $\sim$6\% only for OC members for both criteria (see Table~\ref{lirich_params}). This percentage is higher than in comparison to field stars \citep[about 1-2\%][]{Deepak2019}, even though the authors use a much higher limit (1.80\,dex) to select their Li-rich giants where only three of our stars are above. In addition, the rarity of Li-rich giants is disputed with the new results ($\sim$4\%) of \cite{Frasca2022} who however, use the classical 1.50\,dex limit. This showcases that depending on the definition of what consists a Li-rich giant, different results can be drawn in the literature about their presence. The evolutionary state of the Li-rich giants in this work varies mostly between the RC and the RGB phase. If we consider the Li-rich giants in Table~\ref{lirich_params} from both criteria, the majority of the stars are located at the red clump (13/22), some lie at the (upper) RGB phase (7/22), and for two we could not conclude on their phase. 

Moreover, we have identified stars at the same evolutionary stage in the same cluster with very different Li abundances. We have classified 10 stars with high Li abundance in five different OCs. These stars are divided into the RGB base (2/10), RGB (3/10) or fall at the LB (5/10). If we add these giants to the total amount of Li-rich stars (excluding the stars in the RGB base), the percentage of Li-rich giants in our sample reaches 9-10\%. At this point it is necessary to mention that the limits of defining Li-rich giant stars depend on various assumptions that are difficult to constrain to a single value limit. As proposed in many other works, comparing stars with the same stellar parameters, mass and age can be a clearer indicator to evaluate Li enrichment.

Finally, we find that the projected rotational velocity tends to be higher with Li abundance and with stellar mass. The majority of the clusters with Li-rich giants are young (5/6) and contain massive stars. We have also characterized a Li-rich yellow straggler. 
 
Open clusters are essential laboratories to examine Li evolution because we can easier identify the stellar evolutionary stage and investigate the effect of mass as low mass stars experience Li depletion differently than high mass stars. Taking this into consideration, our work suggests that there is no universal Li production event and that the fraction of Li-rich giants might be higher than previously estimated.

\begin{acknowledgements}

We thank the referee for the very useful comments to improve this paper. This research has been partially supported by the following grants: MIUR Premiale "{\em Gaia}-ESO survey" (PI Sofia Randich), MIUR Premiale "MiTiC: Mining the Cosmos" (PI Bianca Garilli), the ASI-INAF contract 2014-049-R.O: "Realizzazione attivit\`a tecniche/scientifiche presso ASDC" (PI Angelo Antonelli), Fondazione Cassa di Risparmio di Firenze, progetto: "Know the star, know the planet" (PI Elena Pancino), and Progetto Main Stream INAF: "Chemo-dynamics of globular clusters: the {\em Gaia} revolution" (PI Elena Pancino). J.H.C.M. is supported in the form of a work contract funded by Fundação para a Ciência e Tecnologia (FCT) with the reference DL 57/2016/CP1364/CT0007. E.D.M. acknowledges the support by the Investigador FCT contract IF/00849/2015/CP1273/CT0003 and by the Stimulus FCT contract 2021.01294.CEECIND. This work was supported by FCT - Fundação para a Ciência e a Tecnologia through national funds by the following grants: UIDB/04434/2020; UIDP/04434/2020; 2022.04416.PTDC. We thank François Bouchy and Xavier Dumusque for coordinating the shared observations with HARPS and all the observers who helped collecting the data.
\\
This research has made use of the WEBDA database, operated at the Department of Theoretical Physics and Astrophysics of the Masaryk University. This work has made use of the VALD database, operated at Uppsala University, the Institute of Astronomy RAS in Moscow, and the University of Vienna. This work presents results from the European Space Agency (ESA) space mission {\em Gaia}. {\em Gaia} data are being processed by the {\em Gaia} Data Processing and Analysis Consortium (DPAC). Funding for the DPAC is provided by national institutions, in particular the institutions participating in the Gaia MultiLateral Agreement (MLA). The {\em Gaia} mission website is https://www.cosmos.esa.int/gaia. The {\em Gaia} archive website is https://archives.esac.esa.int/gaia.
\\

We used the python packages: astropy \citep[\url{http://www.astropy.org},][]{astropy2013, astropy2018}, numpy \citep[\url{https://numpy.org},][]{2020NumPy-Array}, PyAstronomy \citep[\url{https://pyastronomy.readthedocs.io/en/latest/index.html},][]{pyastronomy2019}, scipy \citep[\url{https://www.scipy.org},][]{2020SciPy-NMeth}, pandas \citep[\url{https://pandas.pydata.org},][]{mckinney-proc-scipy-2010}, and matplotlib \citep[\url{https://matplotlib.org},][]{Hunter:2007}. \\

\end{acknowledgements}

\bibliography{bibliography} 

\appendix

\section{Stellar parameters}

\begin{table*}
\centering
\caption{The sample coordinates, V and G magnitudes and {\em Gaia} ID.}
\label{sample_params}
\begin{tabular}{lcccccc}
\hline\hline
Star name & RA & Dec & V mag & G mag & {\em Gaia} DR2 ID & Notes \\
\hline
IC2714No110   & 11 17 30.575 & --62 42 27.316 & 11.602 & 11.144 & 5336820503972467328 & \\
IC2714No121   & 11 17 31.158 & --62 43 19.506 & 10.836 & 10.308 & 5336820469612710784 & \\
IC2714No126   & 11 17 31.838 & --62 44 42.352 & 10.899 & 10.451 & 5336820297818569088 & \\
IC2714No190   & 11 16 48.904 & --62 43 43.496 & 11.240 & 10.866 & 5336823115312490368 & \\
IC2714No220   & 11 17 35.217 & --62 43 44.723 & 10.805 & 10.556 & 5336820400893232256 & \\
...           & ...          & ...           & ...    & ...    & ...                 & ... \\
\hline
\end{tabular}
\tablefoot{The coordinates and V\,mag are taken from SIMBAD and WEBDA. The G magnitude is taken from {\em Gaia} DR2. Notes indicate non members (MN), planet hosts (PL) or binaries (B). }
\end{table*} 

\begin{table*}
\centering
\caption{The sample parameters derived in this work from our spectroscopic analysis.}
\label{stellar_params_table_spec}
\begin{tabular}{lcccccccc}
\hline\hline
Star name &  T$_{\rm eff}$ &  $\log g$  &  $[Fe/H]$  &  $\upsilon \sin i$  &  S/N  &  Mass$_{spec}$        &  A(Li)  &  A(Li) NLTE  \\
          & (K)            &  (dex)     &  (dex)  &  (km\,s$^{-1}$)        &       &  M$_{\odot}$ &  (dex)  &  (dex)  \\
\hline
IC2714No110    & 4949 $\pm$ 16 & 2.63 $\pm$ 0.04 & --0.08 $\pm$ 0.01 & 4.3 & 109 & 2.55 $\pm$ 0.10 & 0.52 $\pm$ 0.09  & 0.75 \\
IC2714No121    & 4665 $\pm$ 14 & 2.17 $\pm$ 0.05 & --0.09 $\pm$ 0.02 & 2.8 & 70  & 2.70 $\pm$ 0.22 & 0.07 $\pm$ 0.14  & 0.30 \\
IC2714No126    & 4888 $\pm$ 16 & 2.23 $\pm$ 0.05 & --0.06 $\pm$ 0.02 & 6.8 & 116 & 3.08 $\pm$ 0.08 & 1.04 $\pm$ 0.03  & 1.27 \\
IC2714No190    & 4927 $\pm$ 17 & 2.36 $\pm$ 0.05 & --0.05 $\pm$ 0.02 & 5.1 & 112 & 2.86 $\pm$ 0.08 & 0.81 $\pm$ 0.05  & 1.04 \\
IC2714No220    & 4878 $\pm$ 16 & 2.27 $\pm$ 0.05 & --0.09 $\pm$ 0.02 & 4.9 & 105 & 3.15 $\pm$ 0.08 & 0.82 $\pm$ 0.05  & 1.05 \\
...            & ...           & ...             & ...              & ... & ... & ...             & ...              & ...  \\
\hline
\end{tabular}
\tablefoot{The spectroscopic parameters for stars with $\dagger$ are taken from UVES.}
\end{table*} 

\begin{table*}
\centering
\caption{The sample parameters calculated from methods other than spectroscopy. The Li abundances (A(Li)$_{phot}$) are derived from our spectroscopic analysis with T$_{\rm eff, photo}$, $\log g _{trig}$ and spectroscopic $[Fe/H]$ as inputs.}
\label{stellar_params_table_photo}
\begin{tabular}{lccccc}
\hline\hline
Star name   & T$_{\rm eff, photo}$ & $\log g _{trig}$ & Mass$_{phot}$ & A(Li)$_{phot}$ & A(Li)$_{phot}$ NLTE \\
            & (K)                 &  (dex)           & M$_{\odot}$   &  (dex)         &  (dex) \\
\hline
IC2714No110 & 4841 $\pm$ 22 & 2.49 $\pm$ 0.04 & 2.51 $\pm$ 0.11 & 0.31 $\pm$ 0.10  & 0.54   \\
IC2714No121 & 4535 $\pm$ 22 & 2.01 $\pm$ 0.06 & 2.50 $\pm$ 0.16 & --0.19 $\pm$ 0.16  & 0.04   \\
IC2714No126 & 4831 $\pm$ 13 & 2.29 $\pm$ 0.03 & 3.07 $\pm$ 0.09 & 0.95 $\pm$ 0.03  & 1.18   \\
IC2714No190 & 4838 $\pm$ 31 & 2.40 $\pm$ 0.04 & 2.83 $\pm$ 0.11 & 0.66 $\pm$ 0.05  & 0.89   \\
IC2714No220 & 4918 $\pm$ 61 & 2.31 $\pm$ 0.05 & 3.16 $\pm$ 0.08 & 0.88 $\pm$ 0.05  & 1.11    \\
...         & ...           & ...             & ...             & ...              & ...   \\
\hline
\end{tabular}
\tablefoot{Effective temperature (T$_{\rm eff, photo}$) calculated from photometry, surface gravity ($\log g _{trig}$) calculated from PARSEC isochrones with T$_{\rm eff, photo}$ and {\em Gaia} parallaxes, stellar mass (Mass$_{phot}$) calculated from T$_{\rm eff, photo}$, $\log g _{trig}$ and spectroscopic $[Fe/H]$, Li abundances (A(Li)$_{phot}$) derived from FASMA with T$_{\rm eff, photo}$, $\log g _{trig}$ and spectroscopic $[Fe/H]$ as inputs.}
\end{table*} 

\section{Tests on stellar parameters and Li abundances}

\begin{figure*}
  \centering
   \includegraphics[width=0.4\linewidth]{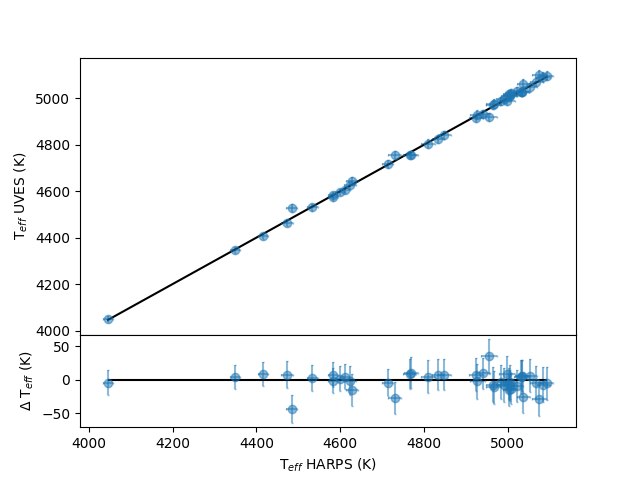}    
   \includegraphics[width=0.4\linewidth]{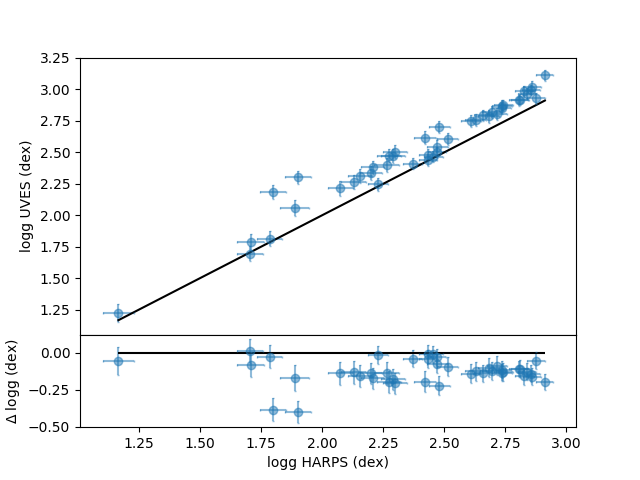} \\    
   \includegraphics[width=0.4\linewidth]{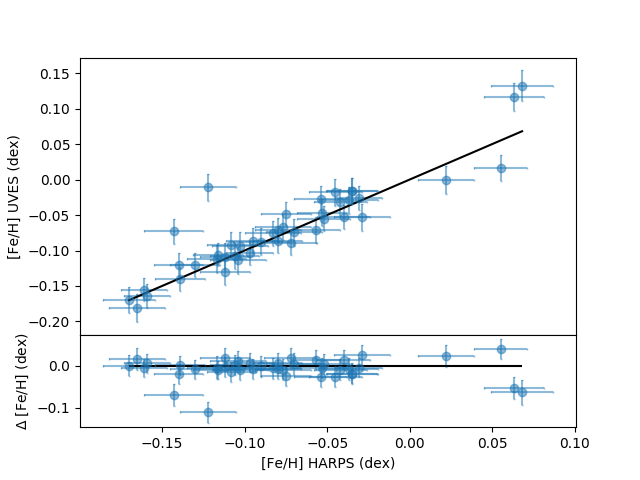}   
   \includegraphics[width=0.4\linewidth]{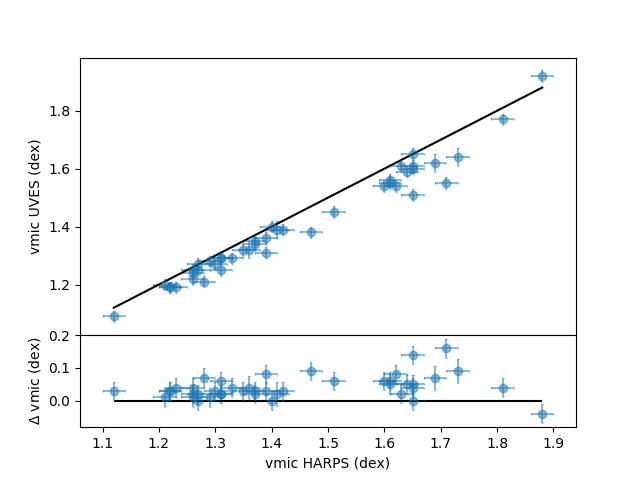} \\
  \caption{Comparison of the stellar parameters derived in this work between HARPS and UVES for 43 stars with spectra available from both spectrographs.}
  \label{param_comparison_spectrographs}
  \end{figure*}

\begin{figure*}
  \centering
   \includegraphics[width=0.4\linewidth]{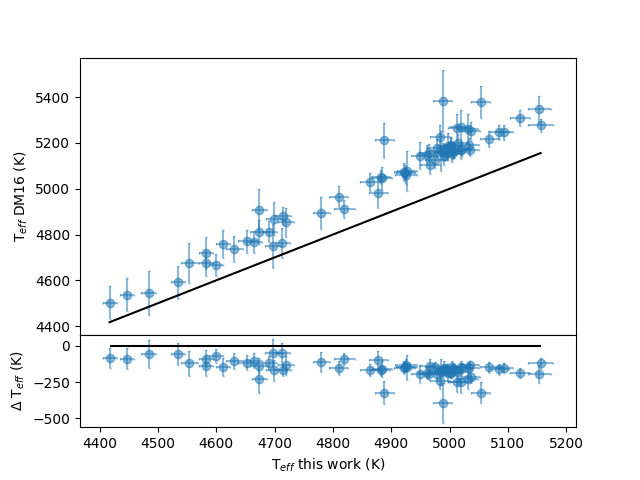}    
   \includegraphics[width=0.4\linewidth]{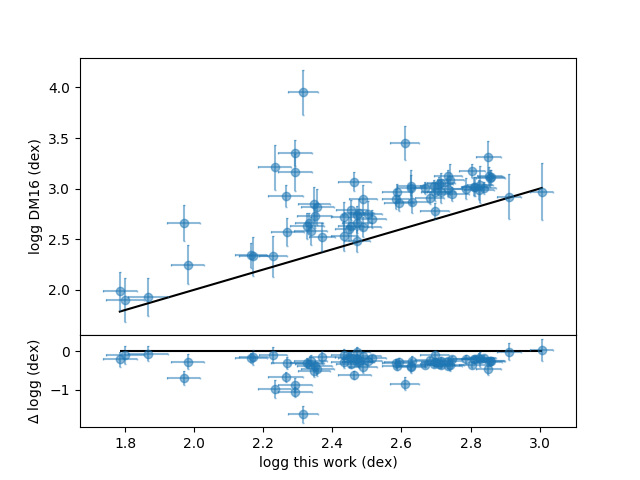} \\    
   \includegraphics[width=0.4\linewidth]{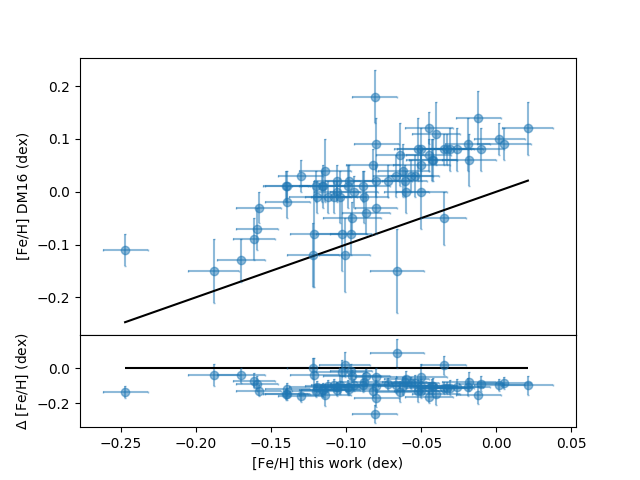}   
   \includegraphics[width=0.4\linewidth]{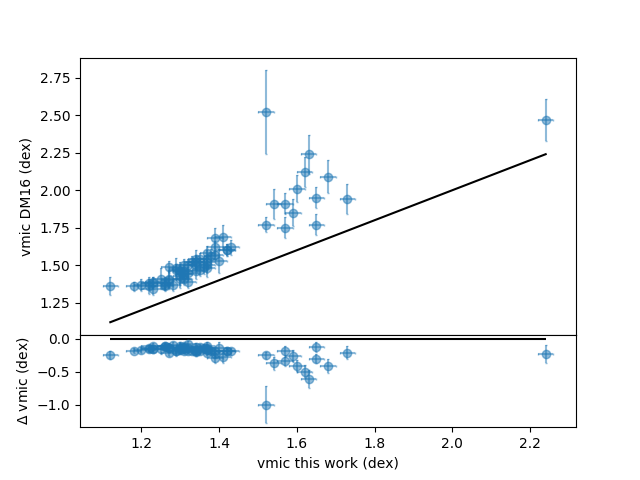} \\
  \caption{Comparison of the stellar parameters derived in this work with our previous results from DM16. The y-axis shows the differences of this work minus DM16.}
  \label{param_comparison_dm16}
  \end{figure*}

\begin{figure*}
  \centering
  \includegraphics[width=0.8\linewidth]{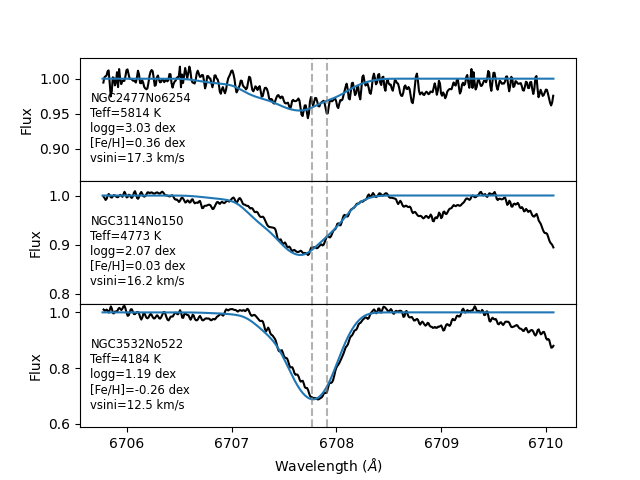} \\
  \caption{The three fastest rotators in our sample. The black lines correspond to the observed spectrum the blue lines to the synthetic fit around the Li lines (dashed gray lines). The minimisation to obtain the best Li abundance is performed in a region of $\pm$1\,$\AA{}$ around the Li lines.}
  \label{high_vsini_fit}
 \end{figure*}

\section{Definition of the maximum A(Li) for OCs}

\begin{figure}
\centering
\includegraphics[width=1\linewidth]{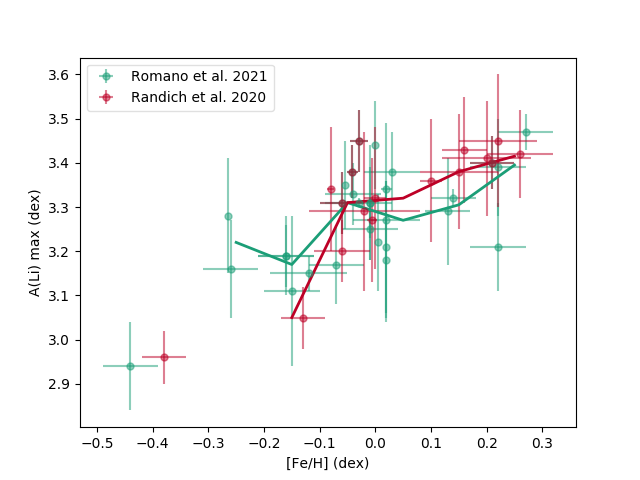} 
\caption{Literature values of the average maximum A(Li) from \cite{Randich2020} and \cite{Romano2021} as a function of $[Fe/H]$ in various OCs from the {\em Gaia}-ESO survey. The lines connect the average A(Li) values of the $[Fe/H]$ bins with size 0.1\,dex.}
\label{limax}
\end{figure}

\section{The HR diagrams}

\begin{figure*}
 \includegraphics[width=0.46\linewidth]{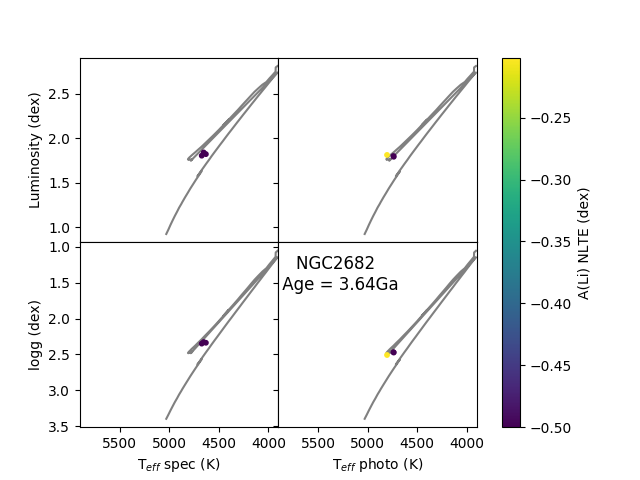}
 \includegraphics[width=0.46\linewidth]{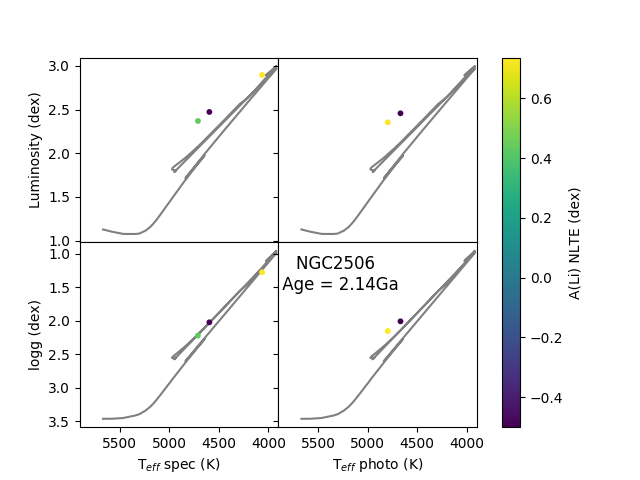} \\
 \includegraphics[width=0.46\linewidth]{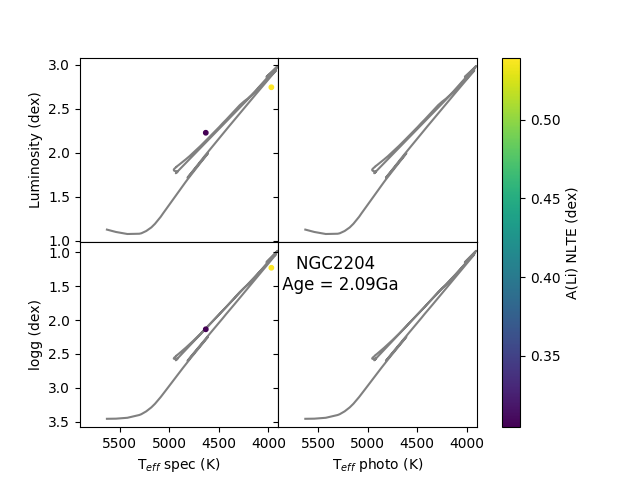}   
 \includegraphics[width=0.46\linewidth]{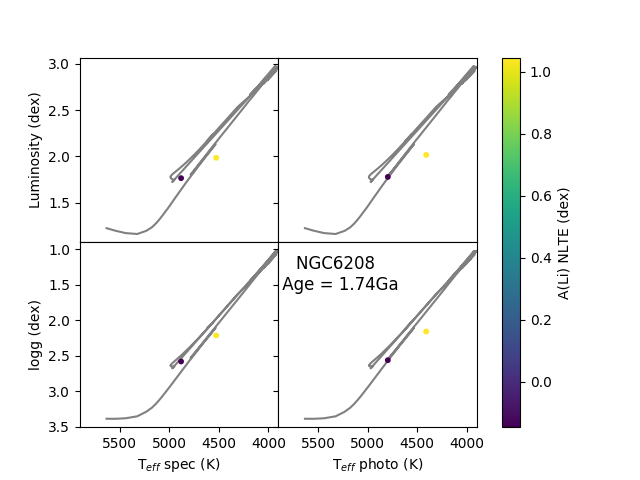} \\
 \includegraphics[width=0.46\linewidth]{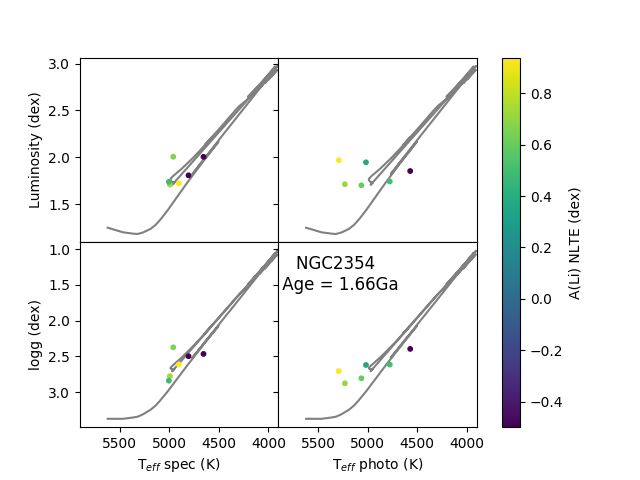} 
 \includegraphics[width=0.46\linewidth]{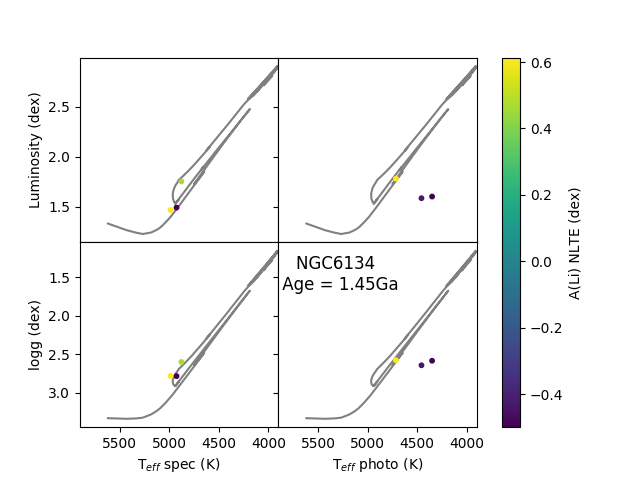} \\
 \includegraphics[width=0.46\linewidth]{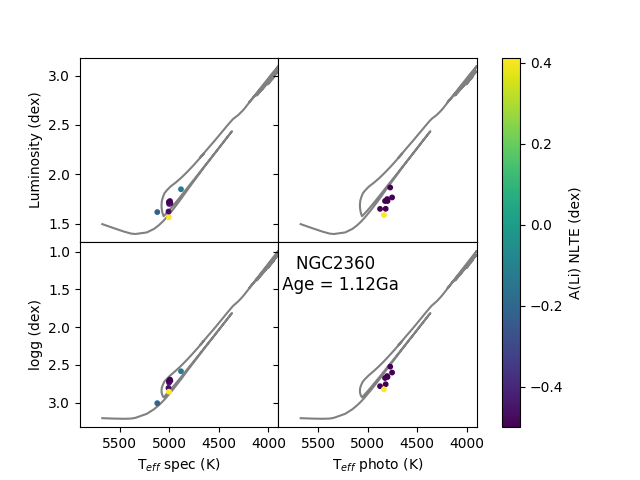}
 \includegraphics[width=0.46\linewidth]{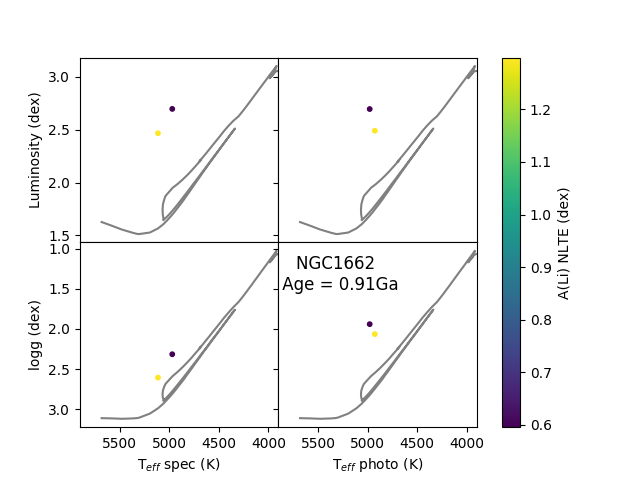} \\
\caption{The HR diagrams of the clusters with N>1 members which do not host Li-rich stars sorted by age. The OC parameters are taken from Table~\ref{cluster_params}. The isochrones are taken from the PARSEC tracks. The left panels are the spectroscopic parameters and in the right the photometric T$_{\rm eff}$ and trigonometric $\log g$. Non members and stars with low signal to noise ratio are excluded.}
\label{iso1}
\end{figure*}

\begin{figure*}
\ContinuedFloat
\includegraphics[width=0.46\linewidth]{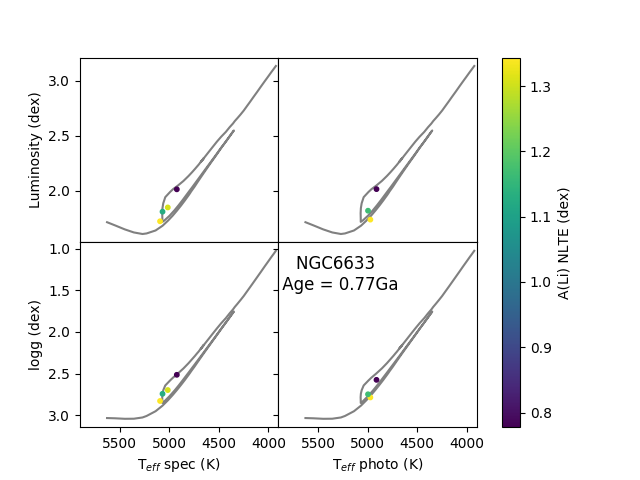} 
 \includegraphics[width=0.46\linewidth]{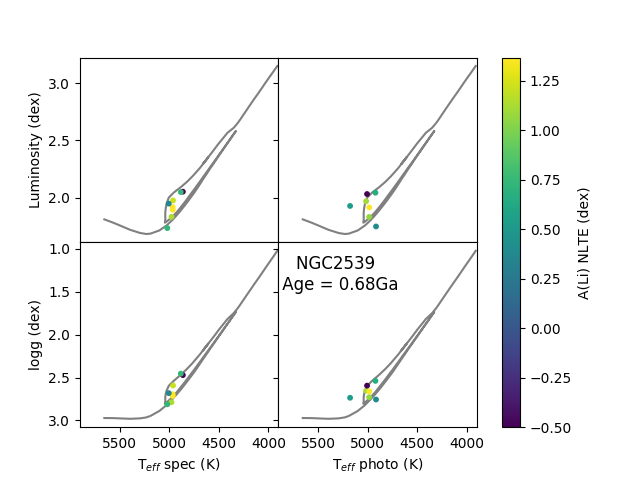} \\
 \includegraphics[width=0.46\linewidth]{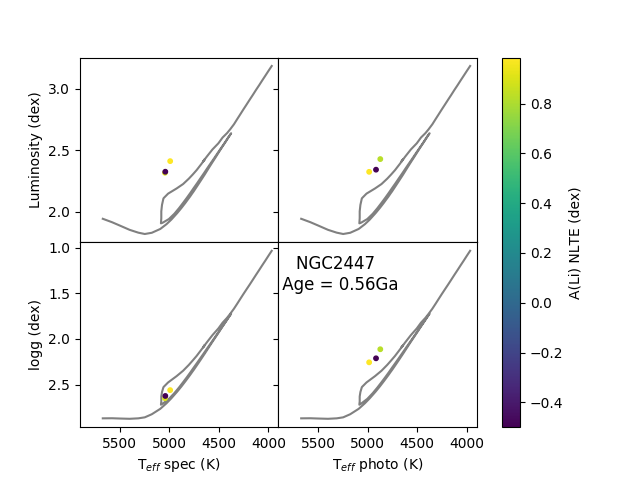} 
 \includegraphics[width=0.46\linewidth]{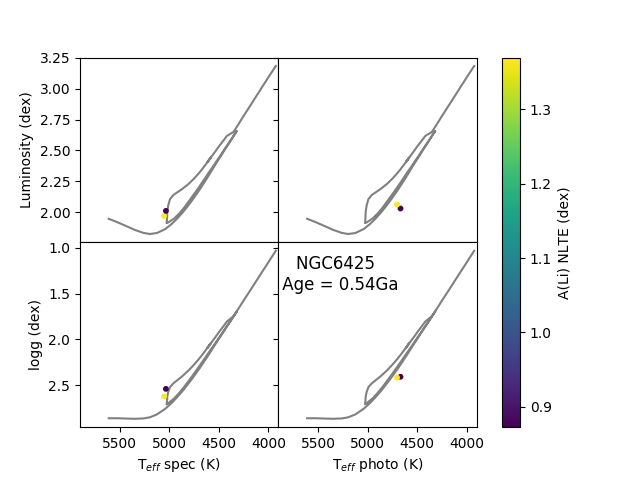} \\
 \includegraphics[width=0.46\linewidth]{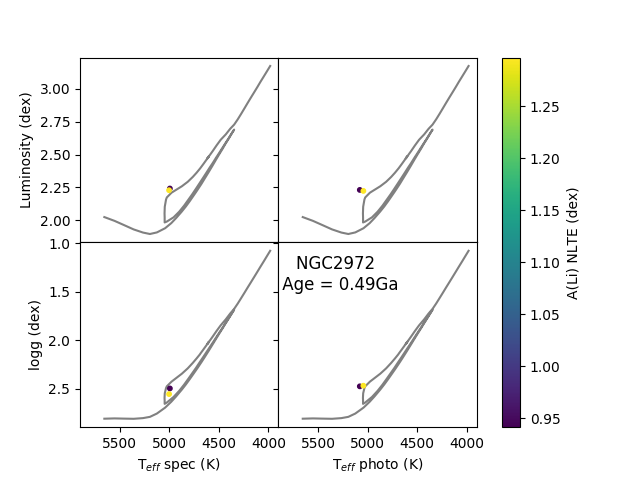} 
 \includegraphics[width=0.46\linewidth]{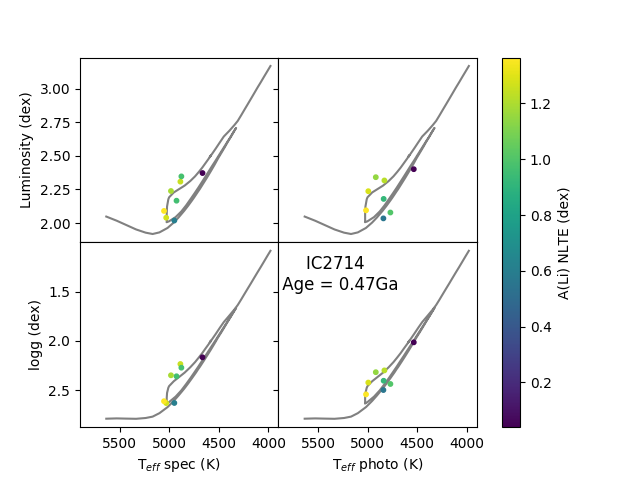} \\
 \includegraphics[width=0.46\linewidth]{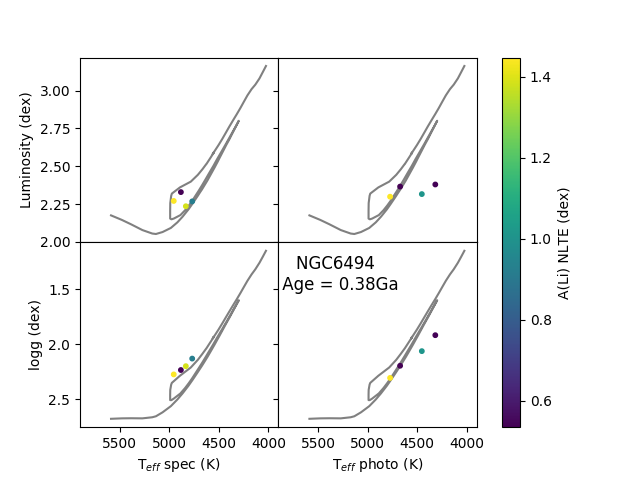} 
 \includegraphics[width=0.46\linewidth]{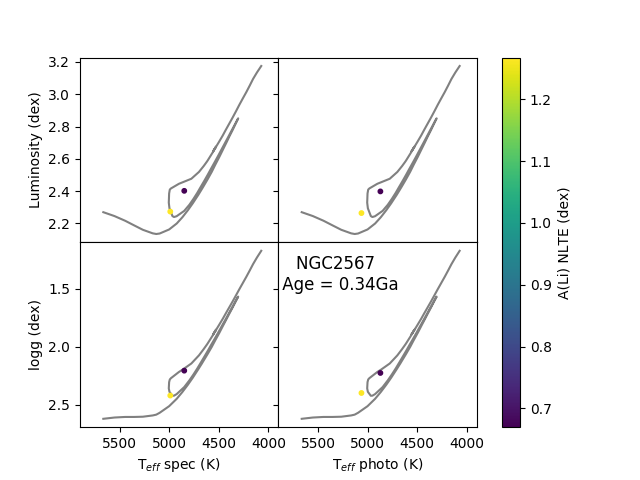} \\
\caption{}\label{iso2}
\end{figure*}

\begin{figure*}
\ContinuedFloat
 \includegraphics[width=0.46\linewidth]{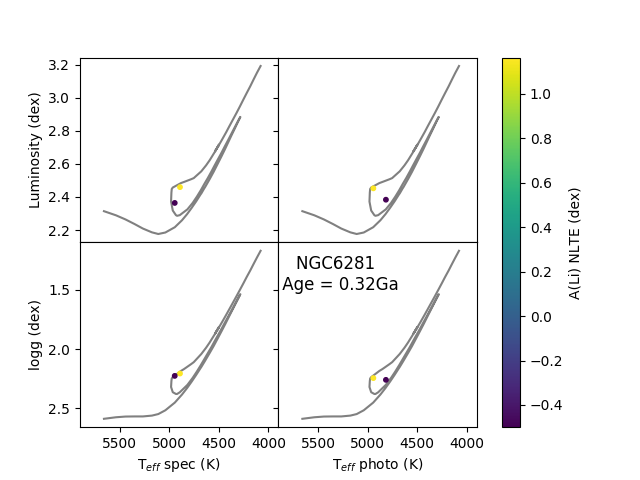} 
 \includegraphics[width=0.46\linewidth]{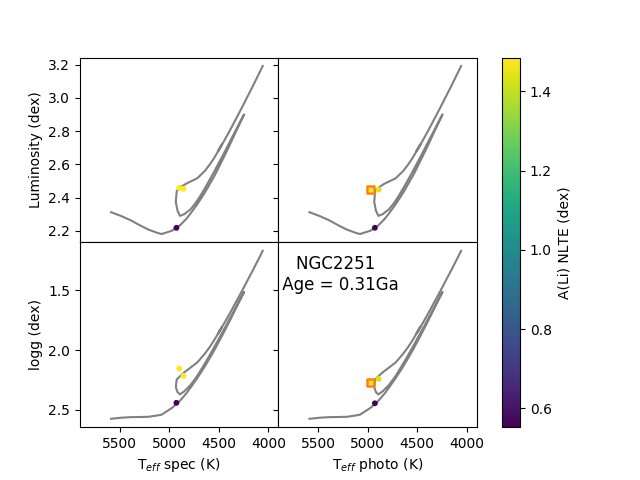} \\
 \includegraphics[width=0.46\linewidth]{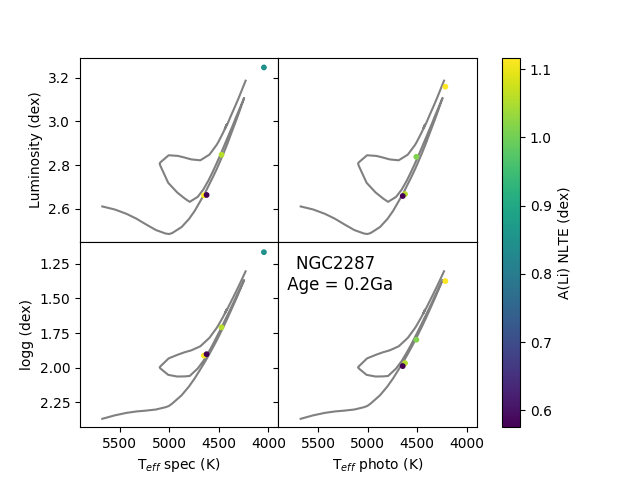} 
 \includegraphics[width=0.46\linewidth]{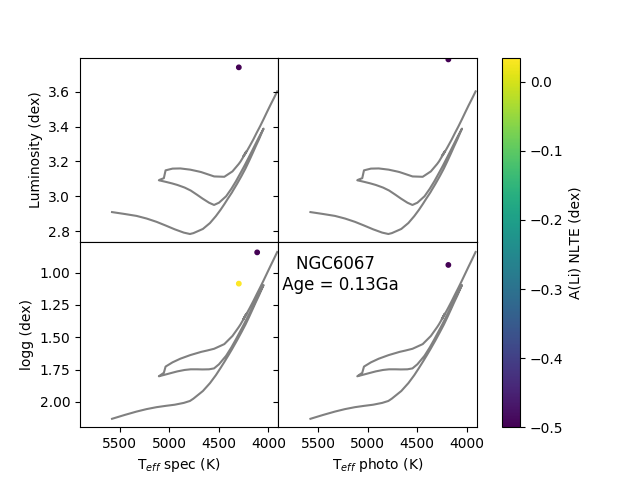} \\
\caption{}\label{iso3}
\end{figure*}
 
\end{document}